\documentclass[aps,prx,twocolumn,superscriptaddress,nofootinbib]{revtex4-2}
\pdfoutput=1
\usepackage{graphicx,amssymb,amsmath,amsfonts,empheq}
\usepackage[dvipsnames]{xcolor}
\usepackage{color}
\usepackage{braket}
\usepackage{latexsym}
\usepackage{upgreek}
\usepackage{dsfont}
\usepackage{bm}
\usepackage{makecell}
\usepackage{multirow}
\usepackage{enumitem}
\usepackage[normalem]{ulem}
\usepackage{url}
\usepackage{hyperref}
\hypersetup{
colorlinks = true,
linkcolor = [rgb]{0.25, 0.41, 0.88}, 
citecolor = [rgb]{0.25, 0.41, 0.88},
urlcolor = [rgb]{0.25, 0.41, 0.88}}
\usepackage{bbm}
\usepackage[utf8]{inputenc}
\usepackage[english]{babel}
\usepackage[T1]{fontenc}
\usepackage{lipsum}
\usepackage{physics}
\usepackage{comment}
\usepackage{wasysym} %

\usepackage{tikz}
\usepackage{tikz-cd}
\usepackage{tkz-base}
\usetikzlibrary{arrows}
\usetikzlibrary{intersections}
\usetikzlibrary{shapes.geometric}
\usetikzlibrary{decorations.pathmorphing, patterns,shapes}
\usetikzlibrary{decorations.markings}

\usepackage{diagbox}

\tikzset{
	partial ellipse/.style args={#1:#2:#3}{
		insert path={+ (#1:#3) arc (#1:#2:#3)}
	}
}

\tikzset{
	mid arrow/.style={postaction={decorate,decoration={
				markings,
				mark=at position .575 with {\arrow[#1]{stealth}}
	}}},
	near arrow/.style={postaction={decorate,decoration={
				markings,
				mark=at position .275 with {\arrow[#1]{stealth}}
	}}},
	far arrow/.style={postaction={decorate,decoration={
				markings,
				mark=at position .800 with {\arrow[#1]{stealth}}
	}}},
}

\pgfdeclarepatternformonly{south west lines}{\pgfqpoint{-0pt}{-0pt}}{\pgfqpoint{3pt}{3pt}}{\pgfqpoint{3pt}{3pt}}{
	\pgfsetlinewidth{0.4pt}
	\pgfpathmoveto{\pgfqpoint{0pt}{0pt}}
	\pgfpathlineto{\pgfqpoint{3pt}{3pt}}
	\pgfpathmoveto{\pgfqpoint{2.8pt}{-.2pt}}
	\pgfpathlineto{\pgfqpoint{3.2pt}{.2pt}}
	\pgfpathmoveto{\pgfqpoint{-.2pt}{2.8pt}}
	\pgfpathlineto{\pgfqpoint{.2pt}{3.2pt}}
	\pgfusepath{stroke}}

\usepackage[dvipsnames]{xcolor}

\renewcommand{\leq}{\leqslant}

\newcommand{\bbZ}{\mathbb{Z}}

\newcommand{\calC}{\mathcal{C}}

\newcommand{\calE}{\mathcal{E}}
\newcommand{\calF}{\mathcal{F}}

\newcommand{\calM}{\mathcal{M}}
\newcommand{\calN}{\mathcal{N}}

\newcommand{\calP}{\mathcal{P}}

\newcommand{\calZ}{\mathcal{Z}}

\newcommand{\sfp}{\mathsf{p}}
\newcommand{\sfq}{\mathsf{q}}

\newcommand{\QM}{\mathrm{QM}}
\newcommand{\QMR}{\mathrm{QMR}}

\newcommand{\bhexagon}{\mathcolor{blue}{\varhexagon}}

\newcommand{\eqnref}[1]{Eq.~\eqref{#1}}
\newcommand{\figref}[1]{Fig.~\ref{#1}}
\newcommand{\appref}[1]{Appendix~\ref{#1}}
\newcommand{\tabref}[1]{Tab.\,\ref{#1}}
\newcommand{\secref}[1]{Sec.~\ref{#1}}

\definecolor{red}{RGB}{240,83,90}
\definecolor{blue}{RGB}{93,123,189}
\definecolor{green}{RGB}{107,189,69}

\definecolor{myred}{RGB}{197,30,58} %
\definecolor{myblue}{RGB}{16,52,166} %
\definecolor{mygreen}{RGB}{79,121,66} %

\definecolor{red1}{RGB}{240,83,90}
\definecolor{blue1}{RGB}{91,98,165}
\definecolor{myturquoise}{RGB}{83,195,189}
\definecolor{lightturquoise}{RGB}{64, 224, 208}
\definecolor{violet}{RGB}{207, 159, 255}
\definecolor{darkturquoise}{RGB}{54, 194, 178}  %
\definecolor{darkviolet}{RGB}{177, 129, 225}    %
\definecolor{canaryyellow}{RGB}{255, 255, 143}
\definecolor{mint}{RGB}{152, 251, 152}
\definecolor{mint2}{RGB}{218, 247, 166}
\definecolor{orange}{RGB}{255,153,28}

\begin{document}

\title{Phases of Floquet code under local decoherence}

\author{Yuchen Tang}
\affiliation{Department of Physics, University of California, Berkeley, CA 94720, USA}
\author{Yimu Bao}
\affiliation{Kavli Institute for Theoretical Physics, University of California, Santa Barbara, CA 93106, USA}

\begin{abstract}
Floquet code is a dynamical quantum memory with a periodically evolving logical space.
As a defining feature, the code exhibits an anyon automorphism after each period, giving rise to a non-trivial evolution of each logical state.
In this paper, we study the Floquet code under local decoherence and perfect measurements and demonstrate that below the decoherence threshold, the code is in a robust phase characterized by the anyon automorphism. 
We first derive the 3D statistical mechanics model for the maximum likelihood decoder of the 2D Floquet code under local Pauli decoherence.
We identify a class of two-qubit Pauli channels under which the 3D statistical mechanics model becomes decoupled 2D models and obtain the threshold for such decoherence channels.
We then propose a diagnostic of the anyon automorphism in the presence of local decoherence.
We analytically show that this diagnostic distinguishes the Floquet code from the toric code under repeated syndrome measurements and undergoes a phase transition at the threshold.
\end{abstract}

\maketitle

\section{Introduction}
The Hastings-Haah Floquet code is a quantum memory with a dynamically evolving logical space generated by a periodic measurement sequence~\cite{hastings2021dynamically}.
The sequence consists of only two-body measurements and is easier to realize in near-term experimental platforms.
Among many variants of dynamical codes that have been proposed~\cite{Vuillot2021, Haah_2022, aasen2022adiabatic, Davydova2023Parent, ellison2023floquetcodestwist, Higgott2024, aasen2023_dead_qubits, zhu2023majorana, Townsend_Teague_2023, Sriram_2023,Zhang2023, motamarri2023symtft,Dua2024, Bauer2024, delafuente2024, fahimniya2024fault, fu2024errorcorrectiondynamicalcodes, Davydova2024, Yan2024, mclauchlan2024, alam2024}, a particular subset called automorphism code, including the Hastings-Haah code, represents a robust many-body dynamics that features anyon automorphism after each period~\cite{aasen2022adiabatic,kesselring2022anyon,Davydova2024}, and non-trivial edge translation, when placed on an open manifold~\cite{aasen2023measurement,sullivan2023floquet}.

The Hastings-Haah code, as well as other automorphism codes, can robustly encode quantum information in the presence of local decoherence~\cite{hastings2021dynamically}.
At any instantaneous time, the logical state is topologically ordered and encodes quantum information non-locally.
The repeated measurements detect the syndromes caused by local decoherence and remove entropy from the system.
This enables possibly non-trivial steady states and allows performing error correction to recover the encoded state based on the measurement results.
This mechanism of preventing entropy accumulation is different from that of the Floquet topological phases in purely unitary evolution, which requires many-body localization or a parametrically long prethermal regime~\cite{Po_2016,Po_2017,Potter2017}.

Specific decoding algorithms in the decohered Floquet codes have been shown to restore encoded information up to finite error thresholds~\cite{gidney2021fault, gidney2022benchmarking, paetznick2023performance, Higgott2024, fahimniya2024fault, delafuente2024}.
The decoding threshold depends on the choice of the algorithm and is upper bounded by the intrinsic threshold of the coherent information, which quantifies the amount of recoverable information from the decohered Floquet evolution.
However, it remains unknown how to determine the intrinsic threshold in the decohered Floquet code and whether the anyon automorphism, as a unique feature of the Floquet code, remains well defined below the threshold.

The intrinsic threshold of topological quantum memory can be derived using two complementary approaches. 
First, for the decoherence that creates incoherent errors, the intrinsic threshold is given by the threshold of the maximum likelihood decoder, which can recover quantum information as long as the coherent information is asymptotically maximum~\cite{hauser2024,niwa2024coherent}.
The maximum likelihood decoding threshold has been derived for the decohered toric code under stabilizer measurements, as well as other topological codes~\cite{dennis2002topological,wang2003confinement,katzgraber2009error,bombin2012strong,kubica2018three,chubb2021statistical,song2022optimal,venn2022coherent,behrends2022surface,behrends2024statistical,bao2024phases}.
The threshold is mapped to the critical point of the two-dimensional random bond Ising model (RBIM) for perfect measurements and the three-dimensional random plaquette gauge theory when measurement readouts are imperfect~\cite{dennis2002topological}.
However, generalizing the maximum likelihood decoding to the Floquet code is not straightforward because the stabilizers are not measured directly and only read out after two consecutive rounds.
Thus, the measurement outcomes of stabilizers depend on errors occurring at multiple time steps, resulting in a 3D statistical-mechanics model with coupling in the time direction even if the measurements are perfect.
This leads to a complicated stat-mech model that is yet to be written down.

Second, one can determine the intrinsic threshold by directly computing the coherent information.
Recently, the coherent information has been computed in topological codes in a single round of decoherence~\cite{Fan2024,lyons2024understanding,lee2024exact,Colmenarez_2024,huang2024,colmenarez2024fundamentalthresholdscomputationalerasure} as well as the decohered toric code under repeated stabilizer measurements~\cite{hauser2024} to probe the intrinsic transition in the decohered density matrix~\cite{Fan2024,bao2023mixed,lee2023quantum,lyons2024understanding,hauser2024,chen2024separability,zhao2024extracting,lee2024exact,sang2024stability,wang2023intrinsic,sohal2024noisy,ellison2024towards,sala2024decoherence,sala2024stability,chen2024unconventional}.
The main technological advance in these works is the stabilizer expansion of the mixed density matrix, which allows expressing the coherent information in terms of defect free energies in statistical mechanics models~\cite{Fan2024}.
However, it is yet to generalize this approach to the case of the Floquet code, which has a dynamically evolving logical space.

In this work, we determine the intrinsic threshold of the decohered Floquet code with perfect measurements and show that the Floquet code phase below the threshold exhibits anyon automorphism characterized by a diagnostic based on quantum relative entropy.
First, we derive the 3D stat-mech model associated with the maximum likelihood decoder.
We also identify a class of two-qubit errors whose syndromes are read out in a single round of measurements.
The related 3D statistical-mechanical model becomes decoupled 2D RBIMs on a honeycomb lattice.
We further analyze the information-theoretical diagnostics that probe the intrinsic transition in the decohered Floquet code.
In particular, we identify a probe of the $e$-$m$ automorphism, based on the quantum relative entropy, in the decohered Floquet code below the threshold.
We map the probe of anyon automorphism and coherent information to statistical mechanics models and show that they undergo the transition simultaneously.
Moreover, this probe distinguishes the Floquet code from the toric code under repeated stabilizer measurements, establishing the Floquet code below the threshold as a separate dynamical phase.

\tableofcontents

\section{Overview}
\begin{table}[t]
    \centering
    \begin{tabular}{|c|c|c|c|}
    \hline 
    & Floquet code & toric code & Trivial \\
    \hline
    $I_c$ & $2\log 2$ & $2\log 2$ & $\leq 0$ \\
    \hline
    $D_{em}^{(n)}$ & $0$ & $\frac{1}{n-1}\log 2$ & $\log 2$ \\
    \hline
    \end{tabular}   
    \caption{Diagnostics of the phases of Floquet quantum memory. The coherent information takes a maximum value $I_c^{max} = 2\log2$ in the encoding phases, while it takes a non-positive value in the trivial phase that does not encode quantum information.
    The Rényi relative entropy $D_{em}^{(n)}$ is zero in the Floquet code phase. In the toric code phase without $e$-$m$ automorphism, $D_{em}^{(n)} =(\log 2)/(n-1)$, and $D_{em} \to \infty$ as $n \to 1$. In the trivial phase, $\rho_1$ becomes maximally mixed and results in $D_{em}^{(n)} = \log 2$. The results are derived for the Floquet code under simple errors, however, are expected to hold for general local Pauli errors.}
    \label{table:diagnostics}
\end{table}

Before proceeding, we briefly summarize the main results of this paper.
We consider the Hastings-Haah code with single edge labels~\cite{gidney2021fault} subject to local decoherence.
For convenience of presentation, we formulate the code on the triangular lattice dual to the honeycomb lattice in most literatures~\cite{hastings2021dynamically,gidney2021fault}.
The vertices and edges of the lattice are colored as in Fig.~\ref{fig:logical_R}.
The qubits live on the triangular plaquettes.
The code involves red (R), green (G) and blue (B) rounds of measurements in sequential order in a period of three rounds.
In each R (G, B) round, measurements are performed on check operators, which are two-body Pauli-X (Y, Z) operators on two qubits adjacent to red (green, blue) edges on the triangular lattice.
The decoherence is described by local quantum channels between two consecutive rounds of measurements.
To identify the intrinsic encoding threshold of the decohered Floquet code, we use two approaches: (1) analyzing the threshold of maximum likelihood decoding; (2) studying the phase transition in the information diagnostics.

\subsection{Maximum likelihood decoding in the Floquet code}
In Sec.~\ref{sec: qec}, we derive the statistical mechanics model that governs the maximum likelihood decoder for the Floquet code under Pauli decoherence.
The maximum likelihood decoder is known to be asymptotically optimal for the toric code under incoherent errors, in the sense that it recovers the encoded state with perfect fidelity as long as the coherent information is asymptotically maximal (up to a correction exponentially small in the system size)~\cite{hauser2024,niwa2024coherent}.

The decoding in the Floquet code bears similarities to that in the toric code.
After each round of measurement, the Floquet code is equivalent to a toric code with a four-dimensional logical space.
The code is stabilized by the measured two-body Pauli operators on the edges and the six-body vertex operators.
We note that the outcome of individual two-body check operators is completely random (with and without Pauli errors), and only vertex operators contain the information of the underlying error operators.
These vertex operators that take the expectation values $-1$ are called \textit{ syndromes}.
We then infer the error configuration by running a decoding algorithm and recover the logical state by applying a recovery operator to remove the syndromes.
The resulting logical state only depends on the homological class of the recovery operator.

The gist of a decoding algorithm is to infer the homological class of the error configuration. 
In particular, the maximum likelihood decoder evaluates the total probability of error strings in the same homological class and determines the class with the highest probability.
The total probability of each class has been shown to map to the partition function of statistical mechanics models in the decohered toric code and other stabilizer codes~\cite{dennis2002topological,wang2003confinement,katzgraber2009error,bombin2012strong,kubica2018three,chubb2021statistical,song2022optimal,venn2022coherent,behrends2022surface,behrends2024statistical,bao2024phases,li2025perturbative}.

The maximum likelihood decoder for the Floquet code has a key difference from that for the toric code.
The stabilizers in the Floquet code given by the six-body vertex operators are only read out after two rounds of measurements. 
Thus, error operators in each round can affect the syndromes observed in multiple rounds, leading to syndrome changes correlated in time.
This results in a 3D statistical mechanics model that is generally coupled in the time direction even if the measurements are perfect, as shown in \secref{sec: qec} and Appendix~\ref{app:single-qubit_stat_mech}.
In contrast, for the decohered toric code under repeated perfect stabilizer measurements, the stat-mech model consists of decoupled 2D random bond Ising models~\cite{dennis2002topological}.

Section~\ref{sec: setup_error_model} introduces a class of two-qubit Pauli errors whose syndromes can be read out after a single round of syndrome measurements, which we call \textit{simple errors}.
For such simple errors, the 3D statistical mechanics model becomes a collection of 2D random bond Ising models (RBIM) on the honeycomb lattice that are decoupled in the time direction.
We obtain the error threshold for such simple errors according to the ferromagnetic transition point of the RBIM.

\subsection{Diagnostics of the Floquet code phase}
Having established the intrinsic threshold of the Floquet code, we propose a probe of the anyon automorphism in the decohered Floquet code in Sec.~\ref{sec:diagnostic}.
We map this probe and the coherent information to statistical mechanics models and show that they undergo a phase transition simultaneously.
This demonstrates the existence of anyon automorphism in the Floquet code below the decoherence threshold.
The probe of anyon automorphism further distinguishes the Floquet code from the toric code under repeated stabilizer measurements.

Section~\ref{sec:diagnostic_em_auto} introduces the probe of e-m anyon automorphism based on the quantum relative entropy.
We consider the evolution of two copies of the Floquet code with different initial states.
We initialize the first copy of the Floquet code in the $+1$-eigenstate of the logical operator $L_m$ that moves the m-anyon along a non-contractible loop. 
We then evolve it through measurements and error channels in an entire period and obtain the final state $\rho_1$.
We initialize the second copy in the maximally mixed state in the logical space and evolve it through a full period of measurements and error channels.
After that, we project the state onto one of the eigenstates of the logical operator $L_e$ that transports the e-anyon along a non-contractible loop, resulting in the final state $\rho_2$. 
The choice of $\pm 1$-eigenstates of $L_e(t)$ depends on the results of the two-body measurements over the entire period of evolution.

We consider the ensemble of trajectories, $\rho_2$ and $\rho_1$, generated by the check measurements in the two copies of the Floquet code and the quantum relative entropy between them,
\begin{equation}
    D_{em}(\rho_2 || \rho_1) :=  \tr \rho_2 \log \rho_2 - \tr \rho_2 \log \rho_1,
    \label{eq:QRE}
\end{equation}
which measures the distinguishability of two density matrices~\cite{vedral2002role}.
Alternatively, one can work with the trajectory $\rho_{1,\mathbf{m}}$ and $\rho_{2,\mathbf{m}}$ with the same check measurement outcome $\mathbf{m}$ and compute their relative entropy $D_{em}$ averaged over trajectories (as detailed in Sec.~\ref{sec:diagnostic_em_auto}).
Intuitively, the logical state of $\rho_{1,2}$ depends on the anyon automorphism, and the relative entropy measures the difference in the logical states of the two.
To facilitate analytic calculations, we consider a specific Rényi version of the quantum relative entropy
\begin{equation}
    D^{(n)}_{em}(\rho_2 || \rho_1) := \frac{1}{1-n} \log \frac{\tr \rho_2 \rho_1^{n-1}}{\tr \rho_2^n},
\end{equation}
which reduces to \eqnref{eq:QRE} in the $n\to 1$ limit.

Section~\ref{sec:diagnostics_stat-mech} maps the R\'enyi-$n$ versions of $D_{em}$ and coherent information $I_c$ in the Floquet code with simple errors to quantities in the statistical mechanics model.
The 3D stat-mech model consists of decoupled 2D $(n-1)$-flavor Ising models in the temporal direction and undergoes a ferromagnetic transition at a finite error rate.
We show that $D_{em}^{(n)}$ and the coherent information $I_c^{(n)}$ probe the ferromagnetic transition in the stat-mech model and therefore exhibit the same threshold.

These information diagnostics distinguish different phases, as summarized in Tab.~\ref{table:diagnostics}.
The coherent information measures the amount of recoverable information from the decohered Floquet code and takes a maximum value $2\log 2$ in the encoding phase, and it is suppressed to a non-positive value in the trivial phase.
The automorphism probe $D_{em}$ can further distinguish the encoding phase of the Floquet code from that of the toric code under repeated stabilizer measurement.
In the Floquet code below the threshold, $\rho_{1,2}$ have identical logical states due to the anyon automorphism, leading to $D_{em}^{(n)} = 0$.
In the toric code, $\rho_1$ and $\rho_2$ are eigenstates of the logical-e and logical-m operators, respectively, giving rise to $D^{(n)}_{em} = (\log 2)/(n-1)$, and $D_{em} \to \infty$ as $n \to 1$.
The values of $D_{em}$ can intuitively be understood as the relative entropy between $\rho_1$ and $\rho_2$ after decoding.
The value of $D_{em}^{(n)}$ is expected to be the same for the Floquet code under other decoherence channels below the threshold. 

We note that the $(n-1)$-flavor Ising model derived for the coherent information is (in the limit $n \to 1$) the Kramers-Wannier dual of the random-bond Ising model for the maximum likelihood decoder in Sec.~\ref{sec: qec}.
Thus, the coherent information exhibits the same threshold $p_c$ as the maximum likelihood decoder, indicating that the maximum likelihood decoder for the Floquet code in Sec.~\ref{sec: qec} is indeed asymptotically optimal.

\section{Setup} \label{sec: setup}
In this section, we introduce the Hastings-Haah Floquet code and specify the error model considered in this work. 

\begin{figure}
    \centering
   \includegraphics[width=0.5\linewidth]{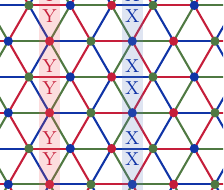}
    \caption{Hastings-Haah code on a 3-colored triangular lattice. The physical qubits are defined on the triangular plaquettes. The logical operators $L_R^Y$ and $L_B^X$ in round R are generated by $E_R^Y$ and $E_B^X$ operators along homologically non-trivial loops. }
    \label{fig:logical_R}
\end{figure}
\subsection{Hastings-Haah code}
We consider the Floquet code on the honeycomb lattice with a single edge label first introduced in Ref.~\cite{gidney2021fault}, which is equivalent to the original Hastings-Haah code (HH code) up to a local basis change~\cite{hastings2021dynamically}.
We take periodic boundary conditions and formulate the code on the triangular lattice as shown in \figref{fig:logical_R} for the convenience of analyzing the maximum likelihood decoder later in this work.

Before specifying the code details, we first assign colors to the vertices and edges in the lattice.
The vertices in the triangular lattice are colorable in three colors red (R), green (G), and blue (B) with adjacent vertices taking different colors.
We assign the third color to the edge connecting the two adjacent vertices of different colors. 
For example, the edge connecting the green and blue vertices is colored red.

The Floquet code involves qubits defined on the triangular plaquettes and is evolved under projective measurements arranged in a periodic pattern.
Each period consists of (R, G, and B) three rounds of measurements in sequential order.
In round $b \in \{R, G, B\}$, we perform measurements on the check operators associated with edges in color $b$.
In particular, we measure the two-qubit Pauli-X (Y, Z) operator on the two adjacent plaquettes of the edges in the color R (G, B).

After an initial full period of measurements (i.e. a warm-up period), the post-measurement state after each round is a stabilizer state with an instantaneous stabilizer group (ISG):
\begin{align} \label{eq:ISG}
    \mathrm{ISG}_R &= \langle E^X_R, V^X_R, V^Y_G, V^Z_B\rangle, \nonumber \\
    \mathrm{ISG}_G &= \langle E^Y_G, V^X_R, V^Y_G, V^Z_B\rangle, \\
    \mathrm{ISG}_B &= \langle E^Z_B, V^X_R, V^Y_G, V^Z_B\rangle, \nonumber
\end{align}
where $V^a_b$ is a six-qubit Pauli operator associated with each vertex.
For example, $V^X_R$ is a product of six Pauli-X operators on the plaquettes surrounding a red vertex.
The vertex operators commute with the measured check operators $E_R^X$, $E_G^Y$, and $E_B^Z$; they belong to all three ISGs, and their values remain unchanged if no error occurs.

\subsection{Logical operators and $e$-$m$ automorphism}\label{sec: setup_logicals}
\begin{figure}
\centering
\includegraphics[width=0.5\linewidth]{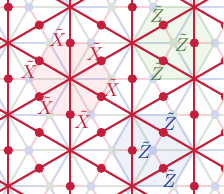}
\caption{Toric code on the R-superlattice after round R measurements. The qubits live on edges in the superlattices. The vertex stabilizers $\Tilde{V}_R^{\Tilde{X}}$ involves $\Tilde{X}$ on the six edges emanating from a red vertex. They correspond to six-body vertex operators $V_R^X$ associated with red vertices on the original lattice.
The plaquette stabilizers $\Tilde{P}_G^{\Tilde{Z}}$ and $\Tilde{P}_B^{\Tilde{Z}}$ are defined as $\Tilde{Z}^{\otimes 3}$ on the edges of the plaquettes on the superlattice. In the original lattice, they are six-qubit vertex operators $V_G^Y$ and $V_{B}^{Z}$ associated with green and blue vertices, respectively.}
\label{fig:superlattices}
\end{figure}

To determine the logical operators, it is convenient to map the Hastings-Haah code at each time step to a toric code on a superlattice.
After each round of measurements, the subspace specified by the measurement outcomes of two-qubit check operators is the same as the Hilbert space of qubits on the edges of a triangular superlattice, as shown in Fig.~\ref{fig:superlattices}.
We can then identify the HH code in this subspace with the toric code on the superlattice.
The logical operators of the HH code stem from those of the toric code.

We explicitly write this mapping in \tabref{tab:superlattice mapping}.
In round $b \in \{R,G,B\}$, two qubits adjacent to a $b$-colored edge $\ell$ on the original lattice are mapped to a single qubit living on the edge that crosses $\ell$ in the $b$-superlattice. 
The measured check operators are mapped to the identity operator on the corresponding superlattice.
The stabilizers $V_R^X$, $V_G^Y$, and $V_B^Z$ in the HH code map to the stabilizers of the toric code in the superlattice following the operator mapping rules in \tabref{tab:superlattice mapping}.
For example, in round R, $V_R^X$ maps to the six-body vertex operator $\Tilde{V}_R^{\Tilde{X}}$ on the superlattice, while $V_G^Y$ ($V_B^Z$) maps to the three-body plaquette operators $\Tilde{P}_G^{\Tilde{Z}}$ ($\Tilde{P}_B^{\Tilde{Z}}$) as shown in \figref{fig:superlattices}, where $\Tilde{\cdot}$ denotes an operator in the superlattice.

\begin{table*}[t]
    \centering
    \begin{tabular}{|c|c|c|c|}
    \hline 
    Round & Operator mapping & Stabilizers \\%
    \hline
    R & \shortstack{$(X \otimes I)_R \cong (I \otimes X)_R \to \Tilde{X}$ \\ $(Y \otimes Y)_R \cong (Z \otimes Z)_R \to \Tilde{Z}$} & $\Tilde{V}_R^{\Tilde{X}}$, $\Tilde{P}_B^{\Tilde{Z}}$, $\Tilde{P}_G^{\Tilde{Z}}$ \\ %
    \hline
    G & \shortstack{$(Y \otimes I)_G \cong (I \otimes Y)_G \to \Tilde{X}$ \\ $(X \otimes X)_G \cong (Z \otimes Z)_G \to \Tilde{Z}$} & $\Tilde{V}_G^{\Tilde{X}}$, $\Tilde{P}_B^{\Tilde{Z}}$, $\Tilde{P}_R^{\Tilde{Z}}$ \\ %
    \hline
    B & \shortstack{$(Z \otimes I)_B \cong (I \otimes Z)_B \to \Tilde{X}$ \\ $(Y \otimes Y)_B \cong (X \otimes X)_B \to \Tilde{Z}$} & $\Tilde{V}_B^{\Tilde{X}}$, $\Tilde{P}_R^{\Tilde{Z}}$, $\Tilde{P}_G^{\Tilde{Z}}$ \\ %
    \hline
    \end{tabular}   
    \caption{Mapping the floquet code to the toric code on the superlattice. In round $b$ with $b \in \{R,G,B\}$, two qubits adjacent to a $b$-colored link is mapped to a single qubit on an edge of the $b$-superlattice. The operators acting on the qubits adjacent to a $b$-colored edge is denoted with $(\cdot)_b$. The superlattice operators are labelled with $\Tilde{\cdot}$ to differentiate from original lattice operators.}%
    \label{tab:superlattice mapping}
\end{table*}

We can write down the logical operators of the HH code according to those in the superlattice toric code.
The logical operators, represented by $L^a_b$, are given by a product of edge operators $E^a_b$ of the same colors and Pauli operators along homologically non-trivial cycles as shown in \tabref{tab:logical_operators} and \figref{fig:logical_R}.
\begin{table}[h]
    \centering
    \begin{tabular}{c c c}
        & e-logical & m-logical \\
        \bf{R}:& $L^Y_R \cong L^Z_R$ & $L^X_B \cong L^X_G$ \\
        \bf{G}:& $L^Y_R \cong L^Y_B$ & $L^X_G \cong L^Z_G$ \\
        \bf{B}:& $L^Y_B \cong L^X_B$ & $L^Z_G \cong L^Z_R$ 
    \end{tabular}
    \caption{Logical operators of each round are string operators that transport $e$($m$)-anyon along non-contractible loops. $e$ and $m$ logicals along two different non-contractible loops anticommute and form logical Pauli-X and Z operators.}
    \label{tab:logical_operators}
\end{table}
The anti-commuting logical operators along two non-contractible loops are called e- and m-logical, $L_e$ and $L_m$, giving rise to the logical Pauli-X and Pauli-Z operators, respectively.
Two logical operators $L_b^a$, related by "$\cong$" in \tabref{tab:logical_operators}, encode the same logical information, as they differ by the measured check operators in each round.
We note that there exists a set of logical operators that encodes quantum information in two consecutive rounds, e.g. $L_R^Y$ and $L_G^X$ in rounds R and G, therefore, the information is preserved during the evolution.

The HH code has an additional defining feature -- $e$-$m$ automorphism after each period.
The logical space undergoes a non-trivial evolution after each period, mapping an eigenstate of the e-logical to that of the m-logical.
Suppose we initialize the code in an eigenstate of the e-logical ($L_R^Y$ and $L_R^Z$) in round R.
The code will evolve into the eigenstate of the e-logical operator $L_B^Y\cong L_B^X$ in round B and further evolve into an eigenstate of an m-logical operator $L_B^X$ in round R of the next period.
We remark that the choice of e and m logical operators in each round is arbitrary, and the e-m automorphism is independent of this choice.

The anyon automorphism can be trivial if we consider an alternative measurement sequence.
For example, instead of measuring check operators $E_B^Z$ in round B, one directly measures the six-body stabilizers $V_R^X$ and $V_G^Y$ in the entire system.
In this case, the e- and m-logicals map to themselves, and the logical state is invariant after each period, similar to the toric code under repeated stabilizer measurements.
We refer to the phase that retains information but does not exhibit $e$-$m$ automorphism as the \emph{toric code phase}.
Later in this work, we will show that the anyon automorphism also distinguishes the decohered Floquet code from the decohered toric code.

One can further envision a transition between the toric code phase and the Floquet code phase by interpolating between the two cases.
More specifically, one can consider omitting each $E_B^Z$ measurement in round B with a probability $q$.
In the case that one or more $E_B^Z$ measurements with a red (green) hexagon are missing, we perform an additional measurement of the six-body operator $V_R^X$ ($V_G^Y$).
When tuning between the Floquet code phase ($q = 0$) and the toric code phase ($q = 1$), there is a bond percolation transition in the configuration of $E_B^Z$ measurements. 
The transition in the anyon automorphism may coincide with the percolation transition, which we leave for a future study.

\subsection{Error model} \label{sec: setup_error_model}
We consider errors in the Floquet code described by the local quantum channel
\begin{equation}
    \rho \to \calN[\rho]
    \label{eq:quantum_channel}
\end{equation}
occurring before measurements in each round.
For example, $\calN = \circ_\triangle \calN_\triangle$ can be the composition of single-qubit error channels with each channel describing bit flip with probability $p$, i.e. $\calN_\triangle [\cdot] = (1-p) [\cdot] + p X_\triangle [\cdot ] X_\triangle$, acting on qubits on the triangular plaquettes.

The error correction of the Floquet code with general local errors is complicated because each error operator causes syndrome changes in multiple rounds.
In this work, we focus mainly on the simple error model, whose syndromes can be read out after one round of measurements, as explained in \secref{sec:qec_detection}.
We consider a total of six types of two-qubit Pauli errors on neighboring qubits given by $E_R^Y$, $E_R^Z$, $E_G^X$, $E_G^Z$, $E_B^X$ and $E_B^Y$\footnote{In principle, we can also include two-qubit errors $E_R^X$, $E_G^Y$, and $E_B^Z$. However, they cannot create logical errors and are trivial to correct in the Floquet code.}.
%
In each round, all six types of errors can occur,
\begin{equation}
    \calN = \calN_R^Y \circ \calN_R^Z \circ \calN_G^X \circ \calN_G^Z \circ \calN_B^X \circ \calN_B^Y \label{eq:RGB_channels}
\end{equation}
where $\calN_b^a = \circ_\ell \calN_{b,\ell}^a$ with $\ell$ denoting the edge associated with the two-qubit channel, $a \in \{X,Y,Z\}$, and the quantum channel for each error type is
\begin{equation}
    \calN_{b,\ell}^a[\rho] = (1-p) \rho + p E_{b,\ell}^a \rho E_{b,\ell}^a.
    \label{eq:errors}
\end{equation}
We set the error rate $p$ to be the same for each type of simple error in the rest of the paper to demonstrate the decoherence-induced transition, although our analysis applies to general error rates of simple errors.

%
%

%
%
%
%
%
%
%
%
%

%
\section{Error correction in Floquet code}\label{sec: qec}
In this section, we describe the error correction in the Floquet code.
We first explain the detection of errors based on the measurement results in the Floquet sequence in \secref{sec:qec_detection}.
We then focus on the simple error model in Eq.~\eqref{eq:RGB_channels}, derive the statistical mechanics model in \secref{sec:qec_maximum_likelihood} that governs the maximum likelihood decoder, and determine the threshold $p_c = 0.0119$. 
Our derivation of the stat-mech model can be generalized to general Pauli errors. We provide an explicit derivation for the single-qubit Pauli-X errors in Appendix~\ref{app:single-qubit_stat_mech}.

\subsection{Error detection and recovery}\label{sec:qec_detection}
In the Floquet code, the decoder infers the underlying errors based on the measurement results of the check operators.
For general Pauli error channels, each individual check operator has a completely random outcome, and the information about error operators is contained in the values of the stabilizers $V_R^X$, $V_G^Y$, and $V_B^Z$ (i.e., the product of check operators), called \emph{error syndromes}.
In what follows, we detail the error detection process. We assume that all measurements have perfect readouts.

\begin{figure}[t]
    \centering
    \includegraphics[width=0.85\linewidth]{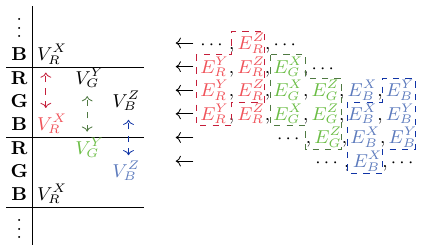}
\caption{Detection of simple errors in the Floquet code. Syndromes of $V_R^X$, $V_G^Y$ and $V_B^Z$ are read out in rounds $B$, $R$, and $G$, respectively. Each type of syndrome change is caused by two types of simple errors in the previous three rounds, enclosed by dashed lines.}
\label{fig:simple_error_detection}
\end{figure}

The vertex stabilizers are given by the product of check operators measured in two consecutive rounds.
For example, the value of $V^X_R = V_R^Y V_R^Z$ defined on a red vertex $v_R$ is inferred from $V_R^Y = \prod_{i \in v_R} Y_i$, a product of $E_G^Y$ measured in round G, and $V_R^Z = \prod_{i \in v_R} Z_i$, a product of $E_B^Z$ measured in round B.
\begin{equation*}
    \begin{tikzpicture}[baseline={([yshift=-.5ex]current bounding box.center)},scale=0.7]
    \node[inner sep=0pt] (pdfimage) at (0,0)
        {\includegraphics[width=0.08\textwidth]{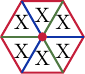}};
    \end{tikzpicture}
    =
    \begin{tikzpicture}[baseline={([yshift=-.5ex]current bounding box.center)},scale=0.7]
    \node[inner sep=0pt] (pdfimage) at (0,0)
        {\includegraphics[width=0.08\textwidth]{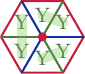}};
    \end{tikzpicture}
    \cdot
    \begin{tikzpicture}[baseline={([yshift=-.5ex]current bounding box.center)},scale=0.7]
    \node[inner sep=0pt] (pdfimage) at (0,0)
        {\includegraphics[width=0.08\textwidth]{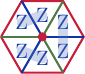}};
    \end{tikzpicture}
\end{equation*}
After one full period, each type of stabilizer is read out once as shown in Fig.~\ref{fig:simple_error_detection}.
By comparing the stabilizers measured in two periods, we obtain the syndrome changes, which contain information about the error operator during the evolution.

Inferring error operators based on syndrome changes is complicated in general because each error operator can create multiple types of syndrome changes that are observed in multiple rounds.
Hence, the syndrome changes are correlated in time, and one has to consider the entire syndrome history to infer errors in a certain round.
This is a key difference from the decoding in the toric code under repeated perfect syndrome measurements, in which syndrome changes are uncorrelated in time, and error operators can be inferred from syndrome changes observed in one round of measurements~\cite{dennis2002topological}.
As a result, the statistical mechanics model for the maximum likelihood decoder for the Floquet code is coupled in the temporal direction, even if the measurements are perfect (as shown for single-qubit Pauli-X errors in \appref{app:single-qubit_stat_mech}).

The decoding in the Floquet code is simple if errors only happen after one round of measurements in each period.
In this case, the observed syndrome changes are only caused by errors that occurred in the same period, and syndrome changes in different periods are uncorrelated.
In the maximum likelihood decoding, this leads to a 3D stat-mech model consisting of 2D models that are decoupled in the temporal direction.

The decoding can also be simple for certain types of errors that occur after every round of measurements.
For the error model introduced in \secref{sec: setup_error_model}, the syndrome changes observed in each period are uncorrelated in time and are only caused by errors between two consecutive syndrome readouts, as shown in Fig.~\ref{fig:simple_error_detection}.
This is because the six error operators in the simple error model in Eq.~\eqref{eq:RGB_channels} only anticommute with one type of stabilizer (e.g. $E_B^X$ and $E_B^Y$ only anticommute with $V_B^Z$).
Hence, each type of syndrome change originates from two types of simple errors and is independent of each other.
We can then perform error correction separately, e.g., correcting $E_B^X$ and $E_B^Y$ based on $V_B^Z$ syndromes only. 
We note that the observed syndrome changes originate from error operators in six different channels in the previous three rounds, as shown in Fig.~\ref{fig:simple_error_detection}.

\begin{figure}
\centering
\includegraphics[width=0.5\linewidth]{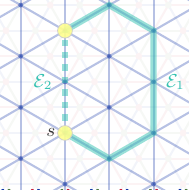}
\caption{Error strings of simple errors $E_B^X$ and $E_B^Y$ on the B-superlattice. The syndrome changes $s$ (highlighted in yellow) live on the vertices of the blue superlattice. 
The solid and dashed lines $\calE_{1,2}$ (highlighted in light blue) represent two equivalent error operators that differ by a product of stabilizers $V_R^X$, $V_G^Y$, and check operators $E_B^Z$.
}
\label{fig:error_recovery_simple}
\end{figure}

To restore the encoded state of the Floquet code, we do not need to apply the recovery operator identical to the error operator.
We take the correction of simple errors $E_B^X$ and $E_B^Y$ as an example.
First, we may apply the recovery operator given by a product of $E_B^X$ operators only.
The potential difference to the error operator by a product of the check operators $E_B^Z$ is unimportant as it cannot create a logical error.
Second, the recovery operator and the error operator can also differ by a product of vertex stabilizers.
This defines equivalent error operators that can be corrected by the same recovery operator.
This can be easily formulated in the B-superlattice, in which the unimportant $E_B^Z$ operator maps to identity as in Tab.~\ref{tab:superlattice mapping}.
The error operator that creates the syndrome $s$ is not unique and maps to a string operator on the superlattice along a path $\calE$ that satisfies $\partial \calE = s$.
Two error operators, $\calE_1$ and $\calE_2$, that create the same syndrome changes differ by a cycle (see \figref{fig:error_recovery_simple}), i.e., $\calE_1 = \calE_2 + C$ with $\partial C = 0$.
They are equivalent and result in the same excited state if the cycle $C$ is trivial.
This defines the equivalence classes of the error operators, according to the homological classes of $\calC$.
The gist of a decoding algorithm is to infer the most likely class of the error string and to apply the recovery operator that belongs to the same class.

\subsection{Maximum likelihood decoder}\label{sec:qec_maximum_likelihood}

The maximum likelihood decoder is a decoding algorithm that is asymptotically optimal for the toric code with Pauli errors.
The maximum likelihood decoder exactly evaluates the probability of the error string in four homology classes $\kappa = 00, 01, 10, 11$ and applies the recovery string in the class with the maximum probability.
In this section, for simplicity, we consider the Floquet code subject to simple errors and map the total probability for each class to the partition function of a statistical mechanics model. 
We provide a derivation that can be generalized to arbitrary local Pauli errors and present a concrete example of the stat-mech model for single-qubit Pauli-X errors in Appendix~\ref{app:single-qubit_stat_mech}.
We emphasize that our analysis of the maximum likelihood decoder assumes that each individual check operator has completely random outcomes in the presence of error channels, and therefore the useful information of the underlying error operator is only in the stabilizer values.
The assumption may not hold for more general quantum channels such as amplitude damping.

We consider the error correction of the simple errors $E_B^X$ and $E_B^Y$ as an example.
For the observed syndrome changes $s$, the probability for error strings in each homology class $\kappa$ is given by
\begin{align}
P_{s, \kappa} = \sum_{\calE; \partial \calE = s, [\calE-\calE_0] = \kappa} P(\calE), \label{eq:class_prob_error_conf}
\end{align}
where $\calE_0$ is a reference string satisfying $\partial \calE_0 = s$.
The fidelity of the maximum likelihood decoder is given by
\begin{align}
\calF = \sum_s P_s \frac{\max_\kappa P_{s,\kappa}}{\sum_\kappa P_{s,\kappa}},
\end{align}
where $P_s = \sum_\kappa P_{s,\kappa}$ is the total probability of observing syndrome $s$.

The total probability for each homology class maps to the partition function of 2D honeycomb-lattice random bond Ising model along the Nishimori line as shown in Appendix~\ref{Appendix:simple_error_model}, i.e.
\begin{align}
    P_{s, \kappa} = \sum_{\{\sigma_i\}} e^{-H_{s,\kappa}}.\label{eq:class_prob_stat_mech}
\end{align}
The Hamiltonian involves Ising spins $\sigma_i = \pm 1$ on the vertices of the honeycomb lattice
\begin{align}
    H_{s,\kappa} = \sum_{\langle i, j\rangle} -J \eta_{ij} \sigma_i \sigma_j, \label{eq:RBIM}
\end{align}
where the bond variable $\eta_{ij}$ takes $-1$ along a reference error string in class $\kappa$ and equals $+1$ everywhere else.
The coupling $J = \log\sqrt{(1-\tilde{p})/\tilde{p}}$ with $\tilde{p} = 6p (1-p)^5 + 20p^3(1-p)^3 + 6p^5(1-p)$.
Here, $\tilde{p}$ is the probability that an odd number of $E_B^X$ and $E_B^Y$ errors occur in this period.

The RBIM undergoes a ferromagnetic-to-paramagnetic phase transition at a finite threshold $p_c$ when increasing the error rate $p$.
The phase transition leads to a sharp change in the decoding fidelity.
In particular, the probability of different homological classes maps to partition functions with different boundary conditions, that is, related by the defect insertion that flips the sign of $\eta_{ij}$ along a non-contractible loop.
Hence, the ratio between two different classes, $P_{s,\kappa}/P_{s,\kappa'} \sim e^{-\Delta F}$, depends on the excess free energy $\Delta F$ of the defect insertion.
In the ferromagentic phase, the defect-free energy scales linearly in the system size, therefore, one homological class has a dominating probability, i.e. $\max_\kappa P_{s,\kappa}/P_s \to 1$ in the thermodynamic limit.
In the paramagnetic phase, the defect-free energy is vanishing; the probability $P_{s,\kappa}$ being the same for each homology class, and the decoding fidelity is suppressed to $1/4$.

The critical point of the honeycomb-lattice RBIM on the Nishimori line has been determined numerically using the transfer matrix method~\cite{Queiroz2006}\footnote{Other works also obtained similar results using Monte Carlo methods~\cite{Takeda2005} and analytical methods such as duality transformation~\cite{Nishimori2006}.}.
The leads to an accuracy threshold 
\begin{equation}
    p_c = 0.0119
\end{equation}
in the simple error model in \secref{sec: setup_error_model}.
We note that if only one of the six simple errors is present, the threshold is $0.0675$.

\section{Diagnostic of Floquet code phase} \label{sec:diagnostic}
\begin{table*}[t]
    \centering
    \begin{tabular}{|c|c|}
    \hline
         $\rho_1$ & $\rho_2$ \\
         \hline
         \shortstack{(1) Initialize in the logical eigenstate of the \\  $m$-logical operator $L_B^X$ in round R\\ (2) Evolve under decohered measurement sequence \\ (G, B, R) for odd periods and end with additional \\G round measurement.
         } & \shortstack{(1) Initialize in the maximally mixed state $\rho_R^0$ \\ in the logical space of round R \\ (2) Evolve under the same measurement sequence as $\rho_1$ \\subject to decoherence \\ (3) Measure $e$-logical operator $L_R^Y$ of round G} \\
          \hline
    \end{tabular}
    \caption{Detailed evolution of $\rho_1$ and $\rho_2$ in the probe of anyon automorphism, $D_{em}$.}
    \label{tab:rho_e_rho_m_procedure}
\end{table*}

In this section, we introduce a probe of $e$-$m$ automorphism in the decohered Floquet code based on quantum relative entropy. 
We develop statistical mechanics models to analytically study the quantum relative entropy and coherent information in the Floquet code under simple error models.
We show that the probe undergoes a phase transition simultaneously with the coherent information and remains constant below the threshold, suggesting that the $e$-$m$ automorphism is a robust property of the Floquet code phase.
Furthermore, the probe of $e$-$m$ automorphism distinguishes the Floquet code from the toric code and trivial phases.
We summarize the values of this probe as well as coherent information in various phases in \tabref{table:diagnostics}.

\subsection{Probe of $e$-$m$ automorphism}\label{sec:diagnostic_em_auto}
The $e$-$m$ automorphism of logical operators after a period of measurements is the defining feature of the Floquet code that distinguishes it from the toric code under repeated stabilizer measurements (as explained in Sec.~\ref{sec: setup_logicals}).
Here, we propose the quantum relative entropy between two states obtained in different evolutions as a probe of anyon automorphism in the Floquet code even in the presence of decoherence.

To formulate this quantity, we consider two copies of the Floquet code after round R measurements.
We initialize the first copy in the eigenstate of the logical operator $L_B^X$, namely an $m$-logical state, and the second copy in the maximally mixed state in the logical space.
Then, we let the two copies of the Floquet code evolve under odd periods of measurements with decoherence and end in round G.
After that, we project the second copy of the Floquet code onto an eigenstate of $L_R^Y$, i.e. an $e$-logical eigenstate. 
Here, the state is renormalized after the projection such that $\tr \rho=1$.
One can regard the projection as a measurement of the logical operator $L_R^Y$ with either post-selection or a feedback to correct the undesired measurement outcome [see Eq.~\eqref{eq:rot_meas} in Appendix~\ref{Appendix: stat-mech diagnostics}].
The evolutions of two copies are summarized in \tabref{tab:rho_e_rho_m_procedure}.
We denote the final states in two copies of the Floquet code as $\rho_1$ and $\rho_2$. 
Here, $\rho_{1(2)} = \sum_{\bf{m}} p_{\bf{m}} \rho_{1(2),\bf{m}} \otimes \ketbra{\bf{m}}$ describes the ensemble of final states generated by the check measurements. 
Each trajectory is labeled by the outcome $\bf{m}$ and its probability $p_{\bf{m}}$.
We note that, during the evolution of Pauli decoherence, $\rho_{1,2}$ only differ in the logical space~\footnote{Consider two logical states under Pauli decoherence in a stabilizer code. The Pauli channels act only on the syndrome space, which is the orthogonal complement of the logical space, and result in two decohered states that differ only in the logical space. 
We note that this is a special feature of the Pauli channel and is not true in general.
}.
Since measuring the check operators does not reveal the logical information, the Born probability $p_{\bf{m}}$ are the same in two copies.

To probe the automorphism, we consider the quantum relative entropy between $\rho_2$ and $\rho_1$
\begin{equation}
    D_{em}(\rho_2 || \rho_1) :=  \tr \rho_2 \log \rho_2 - \tr \rho_2 \log \rho_1.
\end{equation}
In the Floquet code without decoherence, since the $m$-eigenstate evolves to the $e$-eigenstate, $\rho_1$ and $\rho_2$ are identical, leading to a vanishing $D_{em}$.
To facilitate analytic calculations, we instead consider the Rényi relative entropy
\begin{equation}
    D^{(n)}_{em}(\rho_2 || \rho_1) := \frac{1}{1-n} \log \frac{\tr \rho_2 \rho_1^{n-1}}{\tr \rho_2^n},
    \label{eq:Renyi}
\end{equation}
which reduces to $D_{em}$ in the limit $n \to 1$.

We make an important comment that $e$-$m$ automorphism after one full period does not specify a particular $e$-logical eigenstate that an $m$-eigenstate would map to. 
The eigenvalue of $m$-logical operator in the final state $\rho_1$ depends on the check measurement results during the period.
We always project the second copy of the Floquet code onto the $m$-eigenstate that $\rho_1$ would evolve to.

Before proceeding, we make a few remarks.
First, the relative entropy $D_{em}$ can be expressed as the average $D_{em}$ over trajectories, i.e.
\begin{align}
    D_{em}(\rho_2||\rho_1) &= \sum_{\bf{m}} p_{\bf{m}} D_{em}(\rho_{2,\bf{m}}||\rho_{1,\bf{m}}) \nonumber \\
    &= \sum_{\bf{m}} p_{\bf{m}} \tr \rho_{2,\bf{m}} \log \frac{\rho_{2,\bf{m}}}{\rho_{1,\bf{m}}}.
\end{align}
Thus, to compute this quantity, one requires the same outcome for $\rho_{1,2}$.

Second, the number of measurement rounds during the evolution of $\rho_{1,2}$ is in principle arbitrary as long as one projects $\rho_2$ onto the logical state that $\rho_1$ would evolve to in the case without decoherence.
However, the threshold of $D_{em}$ depends on the number of measurement rounds in $\rho_{1(2)}$; the threshold may be higher if the evolution involves only a few rounds.
In this work, we are interested in the behavior of $D_{em}$ when the evolution involves many rounds of measurements.
For the simple error model, this means that we consider greater than or equal to four measurement rounds, for which the threshold of $D_{em}$ is independent of the number of rounds.

Finally, we remark that Ref.~\cite{Vu2024} proposed the expectation value of the logical operator as a probe of
the $e$-$m$ automorphism in the Floquet code with measurement errors.
However, the expectation value takes an exponentially small value in general and is not a good probe in the presence of decoherence.
We further remark that our probe is decoder independent.
To probe the automorphism in real experiments, one can run a specific decoder in the first copy and compare the resulting code state to the desired $e$-logical eigenstate.
However, this may yield a threshold of the automorphism that is suboptimal.

\subsection{Coherent information}\label{sec:diagnostic_coh_info}
Coherent information describes the amount of information retained in the system after decoherence; it distinguishes the encoding from the trivial phase.
To formulate this quantity, we initialize a reference $R$ that is in a maximally entangled Bell state with the logical qubits, which are embedded in the physical Hilbert space of the Floquet code.
We denote the system as $Q$ and the ancilla qubits that store the measurement results as $M$. 
The coherent information is defined as 
\begin{equation}
    I_c(\mathrm{R}, \QM) := S(\rho_{\QM}) - S(\rho_{\QMR})
\end{equation}
where $S(\rho) = -\tr \rho \log \rho$ is the von Neumann entropy~\cite{schumacher1996quantum}.
Computing coherent information analytically is challenging, we consider its Rényi version
\begin{equation}
    I_c^{(n)}(\mathrm{R}, \QM) := \frac{1}{1-n} \log\frac{\tr \rho_{\QM}^n}{\tr \rho_{\QMR}^n}, \label{eq:renyi_coherent info}
\end{equation}
which recovers $I_c$ in the limit $n \to 1$.

In the decohered toric code, it has been shown that $I_c^{(n)}$ is associated with the excess free energy of inserting defects along homologically nontrivial loops in the stat-mech models \cite{Fan2024, hauser2024}.
The stat-mech model undergoes a transition from the encoding phase to a trivial phase detected by $I_c^{(n)}$.
The encoding phase features $I_c^{(n)} = 2\log2$, taking its maximum value.
In Sec.~\ref{sec:diagnostics_stat-mech}, we show that $I_c^{(n)}$ also takes a maximum value $2\log 2$ in the decohered Floquet code below the threshold $p_c^{(n)}$.

\subsection{Statistical mechanics model for diagnostics}
Here, we develop stat-mech models for the information diagnostics in Sec.~\ref{sec:diagnostic_em_auto} and~\ref{sec:diagnostic_coh_info} based on the stabilizer expansion of the decohered density matrix \cite{Fan2024} and obtain the results in \tabref{table:diagnostics}.

\subsubsection{The stabilizer expansion}
The initial state of the Floquet code is the maximally mixed state in the logical subspace $\rho^0_R$ after $E_R^X$ measurements, which can be expressed in terms of the stabilizers $E_R^X$, $V_R^X$, $V_G^Y$, and $V_B^Z$,
\begin{equation}
    \rho^0_R = \frac{1}{4} \prod_\ell \frac{1+E_{R,\ell}^X}{2} \prod_{v_R} \frac{1+V_R^X}{2} \prod_{v_G} \frac{1+V_G^Y}{2} \prod_{v_B} \frac{1+V_B^Z}{2}.
\end{equation}
We analyze the probe of $e$-$m$ automorphism in the stabilizer expansion of the density matrix instead of the error configuration expansion because the former allows studying the dynamical evolution of the logical eigenstates.

Equivalently, $\rho^0_R$ can also be presented in terms of membrane operators $f_R$, $f_G$ and $f_B$ on the original lattice
\begin{equation}
    \rho^0_R = \prod_{\ell} \frac{1+E_{R,\ell}^X}{2} \sum_{f_R} f_R \sum_{f_G} f_G \sum_{f_B} f_B.
    \label{eq:rho_0_original_lattice}
\end{equation}
The summation is over all possible membrane configurations.
We neglect the unimportant overall factor here and note that normalization is needed in calculations.
The membrane operators are defined as
\begin{align}
    f_R = \prod V_R^X, \,
    f_G = \prod V_G^Y, \,
    f_B = \prod V_B^Z,
\end{align}
where $f_b$ consists of six-qubit vertex operators on $b$-colored vertices.
The boundary of membrane operators $f_b$ is a loop along the $b$-color edges of the original lattice.
The $b$-colored vertex operators in the original lattice are plaquette operators on the $b$-honeycomb lattices.
\begin{equation*}
\begin{tikzpicture}[scale=0.5]
\node[inner sep=0pt] (pdfimage) at (0,0)
        {\includegraphics[width=0.5\linewidth]{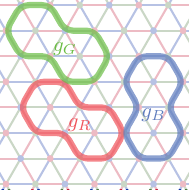}};
\end{tikzpicture}
\end{equation*}

The error channels in \eqnref{eq:RGB_channels} assign real positive weights to the stabilizer according to the loop configuration $g_b \equiv \partial f_b$.
The weight depends on the length $|g_b|$ of the loop on the b-honeycomb lattice, and the error rate plays the role of loop tension.
We provide details of the derivation in \appref{Appendix:corrupted_state_derivation}.

In the actual calculation, we start with the maximally mixed state $\rho = (1/2^N)I^{\otimes N}$ and run a warm-up period of check operator measurements to obtain the Floquet code at $t=0$ as detailed in \appref{Appendix:warm-up}.

\subsubsection{Measurement evolution} \label{Section: evolution}
We now provide details of the measurement evolution leading to the states $\rho_1$ and $\rho_2$ in $D_{em}$ and the states $\rho_{\QM}$ and $\rho_{\QMR}$ in the coherent information.

We model measurements by coupling the system $Q$ to ancilla qubits $M$ and then measuring the ancilla qubit \cite{hauser2024}.
Instead of considering a particular measurement outcome, we consider the ensemble of post-measurement states weighted by their Born probability and with measurement outcomes recorded in the ancilla qubits.
This is equivalent to dephasing (instead of measuring) the ancilla in the measurement basis.
In round $b$, for each measurement of a check operator $E_{b,\ell}^a$ on edge $\ell$, we introduce an ancilla qubit initialized to $\ket{0}_M$.
The ensemble of post-measurement states is obtained by applying a ``measurement'' channel to the initial state
\begin{align}
    \calM_\ell[\cdot] = \calP_{+,\ell} (\cdot) \calP_{+,\ell} \otimes \ketbra{0}_{M} + \calP_{-,\ell} (\cdot) \calP_{-,\ell} \otimes \ketbra{1}_{M},
\end{align}
where the projectors $\calP_{\pm,\ell} = \frac{1 \pm E_{b,\ell}^a}{2}$ are associated with the $\pm 1$ measurement outcomes.
This results in a post-measurement ensemble,
\begin{equation}
\begin{aligned}
    \rho_b(t) &= \circ_{\ell} \calM_{\ell}[\rho(t-1)]\\
    &= \Tilde{\rho}(t-1) \otimes I_M(t) \cdot \prod_\ell \frac{1+E_{b,\ell}^a\otimes Z_{M,\ell}(t)}{2},
\end{aligned}
\label{eq:rho_b}
\end{equation}
where $t$ denotes the round of measurements, the identity operator $I_M(t)$ and Pauli operator $Z_{M,\ell}(t)$ act on the ancilla qubits introduced at time $t$, and $\Tilde{\rho}(t-1)$ is the operator in $\rho(t-1)$ that commute with all measured Pauli operators $E_{b,\ell}^a$.
\emph{Probe of $e$-$m$ automorphism.---}
We now write down the evolution of $\rho_{1,2}$ as in \tabref{tab:rho_e_rho_m_procedure}.
The first copy is initialized in the eigenstate of the $m$-type logical operator $L_B^X$.
Here, we consider a more general setup and prepare the initial state to be a +1 eigenstate of $L_B^X \otimes Z_L$ where $Z_L$ acts on an ancilla qubit,
\begin{align}
\rho_1^0(0) &=  \rho_R^0(1+L_B^X)\otimes \frac{1+Z_L}{2}+\rho_R^0(1-L_B^X)\otimes \frac{1-Z_L}{2}\nonumber \\
&=\rho_R^0 \otimes I_L \cdot (1+L_B^X \otimes Z_L)
\end{align}
This initial state describes the Floquet code in the $L_B^X$ eigenstate, but with the eigenvalue recorded by the ancilla qubit; one can measure the ancilla qubit to initialize the system in the $L_B^X$ eigenstate\footnote{One can also regard this initial state as measuring $L_B^X$ in $\rho_R^0$ and record the measurement outcome in the ancilla qubit.}. 
The second copy of the Floquet code is initialized in $\rho_2^0(0) = \rho_R^0$.
Next, we apply the decoherence channel $\calN$ on $\rho_1^0(0)$ and $\rho_2^0(0)$, 
\begin{align}
    \rho_1^0(0) \to \rho_{R,1}(0), \quad \rho_2^0(0) \to \rho_{R,2}(0).
\end{align}

We then evolve the Floquet codes under three rounds of measurements, $G$, $B$, and $R$.
In each round, we perform the check operator measurements followed by the decoherence channel.
Finally, we measure the $E_G^Y$ check operators and end up in round G at $t = 4$:
\begin{equation}
\begin{aligned}
    \rho_{R,1}(0) & \to \rho_{G,1}(1) \to \rho_{B,1}(2) \to \rho_{R,1}(3) \to \rho_{G,1}^0(4), \\
    \rho_{R,2}(0) &\to \rho_{G,2}(1) \to \rho_{B,2}(2) \to \rho_{R,2}(3) \to \rho_{G,2}^0(4).
\end{aligned}\label{eq:rho12_period}
\end{equation}
We denote the final state of the first copy as $\rho_1 \equiv \rho_{G,1}^0(4)$.
In the second copy, we measure $\rho_{G,2}^0(4)$ in the basis of the $e$-type logical operator $L_R^Y$ and obtain
\begin{equation}
    \rho_{2} =  \rho_{R,2}(4) \otimes I_L \cdot \frac{1+L_R^Y \otimes Z_L}{2}.
\end{equation}
The details of each step of the evolution are given in \appref{Appendix: evolution details}. 

\emph{Coherent information.---}
For concreteness, we consider the coherent information of the decohered measurement sequence for an entire period.
We note that for simple error models, the coherent information does not depend on the number of measurement rounds, as shown in Sec.~\ref{sec:diagnostics_stat-mech}.

The dynamics of $\rho_{\QM}$ starts with the initial state $\rho_R^0$ [\eqnref{eq:rho_0_original_lattice}]. 
The evolution sequence is the same as that in \eqnref{eq:rho12_period} and we define $\rho_{\QM} \equiv \rho_{G,2}^0(4)$.

The dynamics of $\rho_{\QMR}$ starts with a maximally entangled state between the logical qubits of the system $Q$ and the reference qubits (denoted by $R$),
\begin{equation}
    \rho^0_{\QMR,R} = \prod_{l = l_1,l_2} \frac{1+L_{R,l}^Y Z^R_l}{2} \frac{1+L_{B,l}^X X_l^R}{2} \rho^0_R
\end{equation}
where $l=l_{1,2}$ labels the two non-contractible loops on the torus, and $X_l^R$ and $Z_l^R$ are the Pauli-X and Pauli-Z operators of the two reference qubits.
We recall that $L_R^Y$ and $L_B^X$ are the logical Pauli-Z and Pauli-X operators, respectively.

The state $\rho^0_{\QMR,R}$ evolves under decoherence in round R at $t = 0$ followed by the same measurements and decoherence channels as in \eqnref{eq:rho12_period}.
We denote the final state in round G at $t=4$ as $\rho_{\QMR} \equiv \rho_{\QMR,G}^0(4)$.
From time $t=0$ to $t=4$, the logical operators evolve as
\begin{equation}
\begin{aligned}
    L_R^Y &\to L_R^Y \to L_B^Y \to L_B^X \to L_G^X, \\
    L_B^X &\to L_G^X \to L_G^Z \to L_R^Z \to L_R^Y.
\end{aligned}
\label{eq:coherent_info_logical_sequence}
\end{equation}

Finally, we make a few comments regarding the specific evolution we choose for $D_{em}$ and $I_c$.
The initial state of evolution is not limited to $\rho_R^0$ in \eqnref{eq:rho_0_original_lattice} and the evolution does not have to be over a single period. 
For the probe of $e$-$m$ automorphism, the evolution needs to involve an odd number of periods.
For the coherent information, there are no restrictions on where the evolution starts and ends.
In fact, when we are below the threshold $p < p_c$, the information is retained in the memory for a time that is in general exponential in the system size.
\subsubsection{Mapping to the ($n-1$)-flavor Ising model}\label{sec:diagnostics_stat-mech}
Here, we map the $n$-th moment of the density matrix $\tr\rho^n$ in the decohered Floquet code to the partition function of $(n-1)$-flavor Ising model.
For the Floquet code evolved under the decohered measurement sequence, the $n$-th moment reduces to a product of partition functions,
\begin{equation}
    \tr\rho^n = \prod_{b,\tau} \calZ_{b,\tau}. \label{eq:n-moment_stat-mech}
\end{equation}
for each round $b \in \{R,G,B\}$, and period $\tau$.
The partition function is that of the ($n-1$)-flavor Ising model, i.e.  $\calZ_{b,\tau} \equiv \sum_{\{\sigma_{b,\tau}\}} e^{-H_{b,\tau}}$, with
\begin{equation}
    H_{b,\tau} = -3\mu \sum_{\langle i,j \rangle} \left( \sum_{r=1}^{n-1} \sigma_{b,\tau,i}^{(r)} \sigma_{b,\tau,j}^{(r)} + \prod_{r=1}^{n-1} \sigma_{b,\tau,i}^{(r)} \sigma_{b,\tau,j}^{(r)} \right),
\end{equation}
where $\mu = -\log(1-2p)$, Ising spins live on the vertices of a 2D triangular lattice.
The domain wall of Ising spins represents the loop configuration, i.e. $(g_{b,\tau})_{ij} = (1-\sigma_{b,\tau,i}^{(r)} \sigma_{b,\tau,j}^{(r)})/2$ on each edge $\langle i, j\rangle$.
We note that the Floquet code undergoes $T$ periods of evolution, $\tau$ in Eq.~\eqref{eq:n-moment_stat-mech} runs from $1$ to $T+1$, $H_{b,\tau}$ for $\tau = 1,T+1$ has different couplings as detailed in Appendix~\ref{Appendix: stat-mech diagnostics}.

The ($n-1$)-flavor Ising model undergoes a transition at $p_c^{(n)}$ from the paramagnetic phase to the ferromagnetic phase as the error rate increases.
The transition is detected by $D_{em}^{(n)}$ and $I_c^{(n)}$ as summarized in Tab.~\ref{table:diagnostics}.
Moreover, the $(n-1)$-flavor Ising model is dual to the stat-mech model derived for the maximum likelihood decoder and therefore exhibits a threshold $p_c = 0.0119$ when $n \to 1$\footnote{More precisely, the effective Hamiltonian for $n-1$ replicas of the RBIM in Eq.~\eqref{eq:RBIM} is the Kramers-Wannier dual of the $(n-1)$-flavor Ising model. A similar result has been derived for the decohered Toric code~\cite{Fan2024}.}.
This verifies that the maximum likelihood decoder is indeed asymptotically optimal~\cite{hauser2024,niwa2024coherent}.
In what follows, we summarize the result from the mapping and relegate the details to Appendix~\ref{Appendix: Ising mapping} and~\ref{Appendix: stat-mech diagnostics}.

\emph{Coherent information.---}
The coherent information maps to the free energy cost of inserting domain walls along non-contractible loops on the torus,
\begin{align}
    I^{(n)}_c =&  \frac{1}{n-1} \sum_{a=X,Z} \log \bigg( \sum_{\mathbf{d}^a_{l_1}, \mathbf{d}^a_{l_2}} e^{-\Delta F_{\mathbf{d}^a_{l_1}, \mathbf{d}^a_{l_2}}} \bigg)
    \nonumber \\
    & -2\log 2
\end{align}
where $\mathbf{d}_{l_{1,2}}^{X,Z}$ are ($n-1$)-component binary vectors, each component is in $\{0,1\}$ and take the value $1$ if a defect is inserted.
The superscripts $X$ and $Z$ refer to the logical operators that evolve from $L_B^X$ and $L_R^Y$, respectively.
The subscripts $l_{1,2}$ label the two non-contractible loops.
$\Delta F_{\mathbf{d}^a_{l_1}, \mathbf{d}^a_{l_2}}$ denotes the free energy cost of the corresponding domain wall configuration.

In the ($n-1$)-flavor Ising model, the paramagnetic phase corresponds to the encoding phase, with the free energy cost $\Delta F_{\mathbf{d}^a_{l_1}, \mathbf{d}^a_{l_2}} \to 0$, leading to $I^{(n)}_c = 2\log2$.
In the ferromagnetic phase, $\Delta F_{\mathbf{d}^a_{l_1}, \mathbf{d}^a_{l_2}} \to \infty$ unless no defect is inserted.
This leads to $I^{(n)}_c \leq 0$, i.e. loss of quantum information.
\emph{Probe of $e$-$m$ automorphism.---}
The probe of $e$-$m$ automorphism is also associated with the free energy cost of defect insertion along homologically nontrivial loops.
The Rényi relative entropy $D^{(n)}_{em}$ is given by
\begin{equation}
	D_{em}^{(n)}= \frac{1}{1-n} \log{ \bigg( \frac{1}{2^{n-1} } \sum_{\bf{d}}  e^{- \Delta F_{\bf{d}}}  \bigg)}
\end{equation}
where we have defined $\bf{d}$ as a ($n-1$) component binary vector with components $d_r = \{0,1\}$.
If $d_r=1$, a defect is inserted along a homologically non-trivial loop for the $r$-th copy of Ising spins.
In the paramagnetic phase, the defect along the homologically non-trivial loops costs zero free energy in the thermodynamic limit, i.e. $\Delta F_{\bf{d}} \to 0$, leading to $D_{em}^{(n)} = 0$ in the encoding phase of the Floquet code.
On the other hand, in the ferromagnetic phase, the free energy cost of inserting defects along non-contractible loops is linear in system size.
Thus, $e^{-\Delta F_{\bf{d}}} \to 0$, except for the case where $\bf{d} = 0$, which gives rise to $D_{em}^{(n)} = \log 2$. 

If we omit entire rounds of $E_B^Z$ measurements, the dynamical memory is in the toric code phase, which does not have $e$-$m$ automorphism. 
Without decoherence, the state $\rho_1$ remains in the eigenbasis of the $m$-type logical operator $L_G^X$ after the evolution, while $\rho_2$ is in the eigenbasis of the $e$-type logical operator $L_R^Y$.
In the stat-mech model governing $D_{em}^{(n)}$ between decohered toric code states, the defect configuration $\bf d$ is subject to an additional constraint $\sum_r d_r = 0 \mod 2$ and only takes $2^{n-2}$ different values.
In the paramagnetic phase (i.e. the encoding phase of the toric code), we have $D_{em}^{(n)} = (\log 2)/(n-1)$.

We remark that the values of $D_{em}^{(n)}$ in various phases have a simple understanding in terms of the relative entropy between logical qubits after error correction.
In the Floquet code phase, $\rho_{1,2}$ after correction are identical, leading to $D_{em}^{(n)} = 0$.
In the toric code phase, the two-qubit logical state after running error correction in $\rho_{1,2}$ are $\ketbra{+}\otimes(I/2)$ and $\ketbra{0}\otimes(I/2)$ with the first logical qubit in the logical X and Z eigenstates, respectively.
The relative entropy between these two two-qubit states is $D_{em}^{(n)} = (\log2)/(n-1)$.
In the trivial phase, after error correction, $\rho_1$ is maximally mixed in the logical space, i.e. $I^{\otimes 2}/4$, while $\rho_2$ becomes $\ketbra{0}\otimes(I/2)$, leading to $D_{em}^{(n)} = \log 2$.

\section{Discussion}
We have shown that the Floquet code is robust under local decoherence up to a finite threshold.
For a class of simple errors, the 3D stat-mech model for the maximum likelihood decoder is decoupled in the temporal direction and reduces to a stack of 2D RBIM on a honeycomb lattice.
The threshold is given by the critical point of the RBIM.
We further show that the Floquet code below the decoherence threshold exhibits $e$-$m$ anyon automorphism as a defining property.
We introduce a probe of anyon automorphism based on the quantum relative entropy.
For simple errors, we analytically show that the probe distinguishes the Floquet code phase below the threshold from the Toric code and the trivial phases.
Our work opens several directions for future studies.
First, our derivation of the 3D stat-mech model relies on the fact that each check measurement has a completely random outcome, and the information about local errors is only contained in the values of stabilizers.
This is the case when the decoherence is described by general Pauli channels; we derive the stat-mech model for single-qubit bit-flip errors in Appendix~\ref{app:single-qubit_stat_mech}.
However, this property does not hold for the most general decoherence channels.
For example, the amplitude damping channel can favor a certain outcome of check operator measurements, and the optimal (maximum likelihood) decoder in this case needs to rely on the outcomes of all check operators.
Second, in this work, we focus on the maximum likelihood decoding in the decohered Floquet code with perfect measurements.
It is worth studying the decoding threshold in the Floquet code with imperfect measurements in the future.
We remark that a specific type of measurement error, randomly omitting measurements in the Floquet code sequence (or its variants), has been studied in a few recent works~\cite{Vu2024, aitchison2024}. 
The phase transition is shown to be in the universality class of bond percolations.

\begin{acknowledgments}
Y.B. thanks Ehud Altman, Zhehao Dai, and Ruihua Fan for helpful discussions and collaboration on related projects.
Y.B. is supported in part by grant NSF PHY-2309135 and the Gordon and Betty Moore Foundation Grant No. GBMF7392 to the Kavli Institute for Theoretical Physics (KITP).
\end{acknowledgments}

\bibliographystyle{ieeetr}
\bibliography{refs.bib}

\appendix
\onecolumngrid
\section{Details of the maximum likelihood decoding for simple error models} \label{Appendix:simple_error_model}

We detail the derivation of the statistical mechanics model for the maximum likelihood decoding in the Floquet code subject to simple errors.
We explicitly show that the total probability for each homology class $P_{s, \kappa}$ is given by the partition function of the 2D honeycomb-lattice RBIM along the Nishimori line in Eq.~\eqref{eq:class_prob_stat_mech}. 

\begin{figure}
\centering
\includegraphics[width=0.25\linewidth]{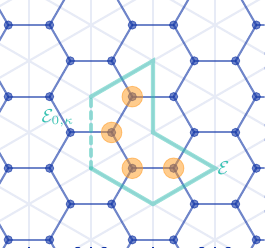}
\caption{Random bond Ising model on the honeycomb lattice. The Ising spins $\sigma_i = \pm 1$ live on the vertices. The domain wall between highlighted spins and the rest represents the cycle $C = \calE_{0,\kappa}+\calE$ with $\calE_{0,\kappa}$ and $\calE$ marked by dashed and solid lines, respectively. The Ising couplings $J\eta_\ell$ on the edges perpendicular to $\calE_{0,\kappa}$ are anti-ferromagnetic, while the rest are ferromagnetic.}
\label{fig:B_honeycomb}
\end{figure}

We take the error correction of the $E_B^X$ and $E_B^Y$ errors as an example.
The total probability $P_{s,\kappa}$ of error strings in each homology class $\kappa$ is given by a summation over equivalent string configurations on the B-superlattice as in Eq.~\eqref{eq:class_prob_error_conf} (illustrated in Fig.~\ref{fig:B_honeycomb})
\begin{align}
    P_{s,\kappa} = \sum_C P(\calE = \calE_{0,\kappa} + C),
\end{align}
where $\calE_{0,\kappa}$ is a reference string in class $\kappa$, and $C$ is a cycle.
Then, the probability of $\calE$ in class $\kappa$ can be written as
\begin{align}
    P(\calE) = P(\calE_{0,\kappa}) P(\calE_{0,\kappa}+C|\calE_{0,\kappa}).\label{eq:prob_E}
\end{align}

First, the probability of the reference error string $\calE_{0,\kappa}$ is
\begin{equation}
    P(\calE_{0,\kappa}) = (1-\Tilde{p})^N \bigg(\frac{\Tilde{p}}{1-\Tilde{p}}\bigg)^{\abs{\calE_{0,\kappa}}},
    \label{eq:prob_E0}
\end{equation}
where $\calE_{0,\kappa}$ is a collection of binary variables on the edges of the B-superlattice, and $\abs{\calE_{0,\kappa}}$ is the length of the error string.
The probability of an error operator occurring on an edge of the B-superlattice is $\Tilde{p} = 6p(1-p)^5 + 20p^3(1-p)^3 + 6p^5(1-p)$.
This is because different error operators can correspond to the same error string on the B-superlattice.
The $E_B^X$ and $E_B^Y$ errors can occur at six time steps in each period, and $\tilde{p}$ describes the probability that they occur for an odd number of times.

Next, the conditional probability is given by
\begin{equation}
    P(\calE_{0,\kappa} + C|\calE_{0,\kappa}) = \bigg( \frac{\Tilde{p}}{1-\Tilde{p}} \bigg)^{\sum_\ell \eta_\ell C_{\ell}},
\end{equation}
where $\eta_\ell = 1 - 2\calE_{0,\kappa,\ell}$, and $C_\ell = 0,1$ is a binary variable with $C_\ell=1$ only if $\ell \in C$.
We can rewrite $P(\calE_{0,\kappa} + C|\calE_{0,\kappa})$ in terms of $u_l = 1-2C_\ell = \pm 1$ and $J = (1/2)\log((1-\Tilde{p})/\Tilde{p})$ as
\begin{equation}
    P(\calE_{0,\kappa} + C|\calE_{0,\kappa}) \propto \prod_\ell e^{J\eta_\ell u_\ell},
    \label{eq:prob_E_final}
\end{equation}
up to an overall factor.
However, the summation over the cycles in terms of \eqnref{eq:prob_E_final} is not easy due to the constraint on $u_\ell$,
\begin{equation}
    \prod_{\ell \in v_B} u_\ell = 1, \label{eq:B-superlattice_cycle_constraint}
\end{equation}
for each blue vertex $v_B$ on the B-superlattice.
It is more convenient to introduce unconstrained variables to express the cycles.

To this end, we introduce Ising spin variables on the vertices of the blue honeycomb lattice and use its Ising domain wall to represent the cycle $C$ (as shown in \figref{fig:B_honeycomb}), i.e.
\begin{align}
    u_\ell = \sigma_i \sigma_j.
\end{align}
where $i$, $j$ on the vertices of blue honeycomb lattice are connected by an edge dual to $\ell$ on the B-superlattice.
In terms of the Ising spins $\sigma_i$, the constraints in Eq.~\eqref{eq:B-superlattice_cycle_constraint} are automatically satisfied.

 \begin{equation}
    P_{s, \kappa} \propto \sum_{\{\sigma_i\}} e^{J \sum_{\langle i, j \rangle} \eta_{ij} \sigma_i\sigma_j},
    \label{eq:RBIM_prob}
\end{equation}
which is the partition function of the RBIM on a honeycomb lattice.
The variable $\eta_{ij} = \pm 1$ and takes the value $\eta_{ij} = -1$ if $\ell \in \calE_{0,\kappa}$.
Moreover, the random bond variable $\eta_{ij} = -1$ with probability $\tilde{p}$, satisfying the Nishimori condition with $e^{-2J} = \tilde{p}/(1-\tilde{p})$~\cite{nishimori1981internal,dennis2002topological}.

\section{Maximum likelihood decoding for single-qubit Pauli errors} \label{app:single-qubit_stat_mech}
In the main text, we have focused on the simple error model in Eq.~\eqref{eq:RGB_channels}, which results in uncorrelated syndrome changes in time. 
In this appendix, we develop the statistical mechanical model of the maximum likelihood decoder for the Floquet code subject to single-qubit Pauli-X errors. 
Our derivation can be generalized to general local Pauli errors.
The general Pauli errors create syndrome changes that are correlated in time and result in a 3D stat-mech model coupled in the temporal direction as demonstrated in the example of Pauli-X errors.

We consider single-qubit bit-flip channels acting on each qubit (on triangular plaquettes in Fig.~\ref{fig:logical_R}) between two rounds of measurements, described by $\calN_X = \circ_\triangle \calN_{X,\triangle}$,
\begin{equation}
    \calN_{X,\triangle}[\rho] = (1-p)\rho + p X_\triangle \rho X_\triangle,
\end{equation}
where $p$ is the error rate.
The bit-flips occur in round R commute with the follow-up $E_R^X$ measurements and are equivalent to those that occur in round G.
For simplicity, we assume that the channel only act before round G and B measurements.
We also assume that the bit-flip channel in each step has the same error rate $p$.

\begin{table}[h]
\begin{tikzpicture}
\node[inner sep=0pt] (pdfimage) at (0,0)
        {\includegraphics[width=0.6\linewidth]{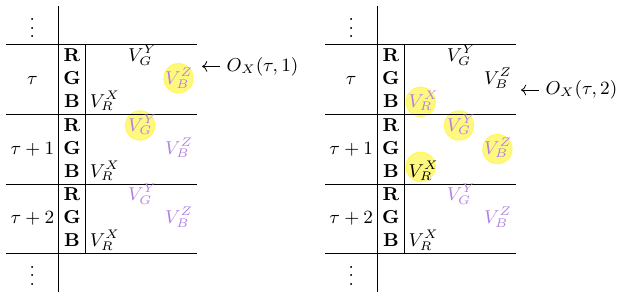}};
\end{tikzpicture}
    \caption{The stabilizers read out in each round of measurements. Consider an error operator occurring after round R (left) or round G (right). The stabilizers affected by the error operator are marked in purple, and syndrome changes are highlighted in yellow.}
    \label{tab:syndrome}
\end{table}

Here, we detail the maximum likelihood decoding of bit-flip errors based on syndrome changes. 
We denote the error operator in round G (B) before $E_G^Y$ ($E_B^Z$) measurements in period $\tau$ as $O_X(\tau,1)$ ($O_X(\tau,2)$).
The error operators can affect the check operator measurements and create syndrome changes in the next few rounds as shown in Tab.~\ref{tab:syndrome}.
We note that to recover the encoded state, we do not need to apply a recovery operator identical to the error operator.
We only need to infer the flipped check operators and apply a Pauli-X string operator that flips a set of check operators that is homologically equivalent to the ones flipped by the error operator.

\begin{figure}[t]
\centering
\includegraphics[width=0.25\linewidth]{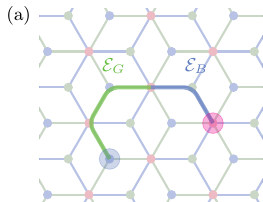}
\includegraphics[width=0.25\linewidth]{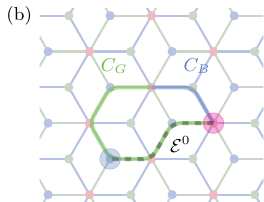}
\includegraphics[width=0.25\linewidth]{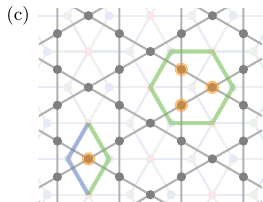}
\caption{(a) An example of the flipped check operators $\calE(\tau) = \calE_G(\tau) + \calE_B(\tau)$ that are compatible with the syndrome changes (highlighted in blue and red). The flipped check operators $\calE(\tau)$ involve two parts on the blue and green edges. (b) A reference string $\calE^0(\tau)$ and deformations $C(\tau) = C_G(\tau) + C_B(\tau)$ that give the flipped check operators in panel (a). (c) Ising spin model on the kagome lattice. Anti-aligned spins (marked in orange) result in domain walls on the original triangular lattice along the green and blue edges, which represent the deformation $C(\tau)$.}
\label{fig:single_qubit_flipped_checks}
\end{figure}

The flipped check operators in round G and B of period $\tau$ are represented by open string variables $\calE_{G}(\tau)$ and $\calE_B(\tau)$ in the triangular lattice, respectively (as shown in Fig.~\ref{fig:single_qubit_flipped_checks}).
The flipped check operators need to be compatible with the syndrome changes and the underlying error operators.
Let $s_R(\tau)$, $s_G(\tau)$ and $s_B(\tau)$ be the syndrome changes of $V_R^X$, $V_G^Y$ and $V_B^Z$ in period $\tau$.
The syndrome changes and the flipped checks satisfy the following relation
\begin{align}
    s_G(\tau+1) = \partial \calE_B(\tau)|_G, \quad s_B(\tau) = \partial \calE_G(\tau)|_B, \quad s_R(\tau) = \partial \calE_G(\tau)|_R + \partial \calE_B(\tau)|_R,
\end{align}
The boundary of $\calE_{G}$ ($\calE_{B}$) includes blue and red (green and red) vertices denoted as $\partial\calE_{G}|_B$ and $\partial\calE_{G}|_R$ ($\partial\calE_{B}|_G$ and $\partial\calE_{B}|_R$), respectively.
We note that the configurations $(\calE_G, \calE_B)$ of the flipped checks that result in the same syndrome changes are not unique; they can be parametrized in terms of the deformation $C_{G(B)}$ from reference configurations $\calE_{G(B)}^0$, i.e.
\begin{align}
    \calE_G(\tau) = \calE_G^0(\tau) + C_G(\tau), \quad \calE_B(\tau) = \calE_B^0(\tau) + C_B(\tau).
\end{align}
Here, the deformations of the reference strings, $C_{G}$ and $C_B$, are in general open strings that must satisfy
\begin{align}
    \partial C_G(\tau) + \partial C_B(\tau) = 0, \quad \partial C_G(\tau), \partial C_B(\tau) \in v_R,
\end{align}
where we recall that $v_R$ is the collection of red vertices.
We emphasize that $C_G(\tau)$ and $C_B(\tau)$ in the above equation belong to the same period $\tau$.
$C_G(\tau)$ can be an open string that ends on red vertices because the red stabilizers $V_R^X$ are not read out in round G, while $C_G(\tau) + C_B(\tau)$ has to be closed because the syndromes associated with $V_R^X$ and $V_G^Y$ are obtained in the next two rounds.
In what follows, we introduce string configurations on both green and blue edges in the triangular lattice
\begin{align}
    \calE(\tau) = \calE_G(\tau) + \calE_B(\tau), \quad \calE^0(\tau) = \calE_G^0(\tau) + \calE_B^0(\tau), \quad C(\tau) = C_G(\tau) + C_B(\tau),
\end{align}
to represent the configuration of the flipped checks, the reference configuration, and closed-loop deformations in each period $\tau$, respectively.
In the case where $C(\tau)$ contains flipped checks along homologically nontrivial loops, we obtain $\calE(\tau)$ which belongs to a different homological class than $\calE^0(\tau)$.

In the maximum likelihood decoder, a central quantity is the total probability of error strings that create homologically equivalent flipped checks in class $\bm\kappa$
\begin{align}
    P_{s,\bm{\kappa}} = \sum_{\calE|s,\bm{\kappa}} \sum_{O_X} P(\calE|O_X) P(O_X)
\end{align}
Here, since the error operators $O_X$ can flip check operators in multiple rounds, the flipped check operators have correlations in time.
We note that $\bm{\kappa} = (\{\kappa_\tau\})$ is a $T$-component vector that labels the homological class of flipped check operators $\calE(\tau)$ in $T$ periods.
To perform the error correction, one computes 
\begin{align}
    P_{s,\kappa} = \sum_{\bm{\kappa}} P_{s,\bm{\kappa}} \delta_{\kappa = \text{sum}(\bm{\kappa})},
\end{align}
which is the total probability for the overall class $\kappa$.
We then apply a recovery string from the class with the maximum probability to restore the encoded state.

We express the summation over $C$ in $P_{s,\kappa}$ as a summation over unconstrained spin variables.
Specifically, we introduce $\sigma_{i,\tau}$ at the center of each red edge on the triangular lattice.
The locations of these spins are the vertices of a kagome lattice.
We also introduce variables $\eta_{ij,\tau}= \pm 1$ on the edge $\langle i, j\rangle$ of the kagome lattice.
\begin{gather}
    \calE_{ij}^0(\tau) = \frac{1 - \eta_{ij,\tau}}{2}, \quad C_{ij}(\tau) = \frac{1 -\sigma_{i,\tau}\sigma_{j,\tau}}{2}.
\end{gather}
In this way, the flipped check operators $\calE(\tau)$ (in the same homological class as $\calE^0(\tau)$) are represented as
\begin{align}
\calE_{ij}(\tau) = \frac{1 - \eta_{ij,\tau}\sigma_{i,\tau}\sigma_{j,\tau}}{2},
\end{align}

The conditional probability $P(\calE|O_X)$ is a constraint on the error configuration.
Let $x_\sfp(\tau,1), x_\sfp(\tau,2) = \pm 1$ denote the error configuration on plaquette $\sfp$.
The flipped check operator is compatible with the error configuration if
\begin{gather}
    \eta_{ij,\tau}\sigma_{i,\tau}\sigma_{j,\tau} = x_\sfp(\tau-1,2)x_\sfq(\tau-1,2)x_\sfp(\tau,1)x_\sfq(\tau,1), \quad \text{if} \langle i, j\rangle \in e_G,\\
    \eta_{i'j',\tau}\sigma_{i',\tau}\sigma_{j',\tau} = x_{\sfp'}(\tau,1)x_{\sfq'}(\tau,1)x_{\sfp'}(\tau,2)x_{\sfq'}(\tau,2),\quad \text{if} \langle i', j'\rangle \in e_B.
\end{gather}
Here, we abuse the notation and use $\langle i, j\rangle$ ($\langle i', j'\rangle$) to represent the green (blue) edge on the triangular lattice dual to $\langle i, j\rangle$ ($\langle i', j'\rangle$) on the kagome lattice.
We use $\sfp, \sfq$ ($\sfp', \sfq'$) to represent the two plaquettes adjacent to the green (blue) edge $\langle i,j \rangle \in e_G$ ($\langle i',j'\rangle \in e_B$) on the triangular lattice.

We can thus express $P_{s,\bm{\kappa}}$ as
\begin{align}
P_{s,\bm{\kappa}} = \sum_{\{\sigma\}} \sum_{\{x_\sfp\}} \prod_{\tau} &\prod_{\langle i, j\rangle \in e_G} \frac{1 + \eta_{ij,\tau}\sigma_{i,\tau}\sigma_{j,\tau} x_\sfp(\tau-1,2)x_\sfq(\tau-1,2)x_\sfp(\tau,1)x_\sfq(\tau,1)}{2} \nonumber \\
&\prod_{\langle i', j'\rangle \in e_B}\frac{1 + \eta_{i'j',\tau}\sigma_{i',\tau}\sigma_{j',\tau} x_{\sfp'}(\tau,1)x_{\sfq'}(\tau,1)x_{\sfp'}(\tau,2)x_{\sfq'}(\tau,2)}{2} \prod_{\sfp} e^{h x_\sfp(\tau,1) + h x_\sfp(\tau,2)}.
\end{align}
where $h = (1/2)\log((1-p)/p)$, and we have ignored an unimportant overall factor.

Summing over all $x_\sfp$ leads to a stat-mech model for $\sigma_{i,\tau}$. 
The stat-mech model has $\bbZ_2^{\otimes T}$ Ising symmetries that flip the $\sigma_{i,\tau}$ spins at each individual time step $\tau$.
The symmetry is the same as that in the stat-mech models for the Floquet code with simple errors in Eq.~\eqref{eq:RBIM} as well as the toric code under repeated perfect measurements~\cite{dennis2002topological}.
Crucially, the symmetry different from the gauge symmetry in the 3D stat-mech model for the toric code under imperfect measurements.
Moreover, the stat-mech model exhibits coupling in the temporal direction, although the measurements are perfect.
This is different from the case of simple errors and the toric code under repeated perfect measurements, in which the spins at different time steps in the stat-mech model are uncorrelated.

Although we do not perform an explicit calculation here, we do expect a phase transition in the stat-mech model when tuning the error rate.
In the limit as $p = 0$, there are no errors, which means that $x_\sfp = 1$ and $\eta_{ij} = 1$ everywhere.
Thus, Ising spins are perfectly aligned and exhibit ferromagnetic order.
In the limit $p=1/2$, all $x_\sfp$ configurations occur with the same probability.
Thus, the $\sigma$ spins can be in any configuration with equal probability weight.
When reducing the error rate $p$, we expect the stat-mech model of $\sigma$ spins to spontaneously break the $\bbZ_2$ symmetry at a finite error rate $p_{c,single}$.

\section{Details of statistical mapping of diagnostics}
In this section, we provide the details of the statistical mapping of the diagnostics.
\appref{Appendix:warm-up} explicitly writes down the warm-up period that leads to the initial state.
\appref{Appendix:corrupted_state_derivation} derives the stabilizer expansion of the decohered state.
In \appref{Appendix: evolution details}, we start with the initial state obtained after the warm-up period and run the measurement sequence with decoherence in \eqnref{eq:rho12_period}.
We explicitly write down the evolution of quantum states involved in the definition of the $e$-$m$ automorphism probe and the coherent information.
\appref{Appendix: Ising mapping} maps the $n$-th moment of the decohered density matrices obtained after measurement sequence to the partition function of the $(n-1)$-flavor Ising model.
Finally, \appref{Appendix: stat-mech diagnostics} maps the probe of e-m automorphism and the coherent information to physical quantities in the statistical mechanical model, which undergo transitions at the same critical point.

\subsection{Warm-up period} \label{Appendix:warm-up}
Here, we explain the warm-up period of measurements that prepares the maximally mixed state in the logical space of the Floquet code at $t = 0$.
The warm-up period involves four rounds of measurements.

We start with the maximally mixed state of $N$ physical qubits at $t=-4$,
\begin{equation}
    \rho(-4) = \frac{1}{2^N} I^{\otimes N}.
\end{equation}

First, we measure the check operators in round R by coupling the system to ancillae as in Eq.~\eqref{eq:rho_b}, giving rise to the state at time $t=-3$,
\begin{equation}
    \rho_R^0(-3) = \frac{1}{2^N} I^{\otimes N} \cdot \prod_{\ell_R} \frac{1+E_{R,\ell_R}^X \otimes Z_{M,\ell_R}(-3)}{2}
\end{equation}
where $\ell_R$ is the location of the $E_{R,\ell_R}^X$ check operator.

Next, we measure the check operators $E_{G,\ell_G}^Y$, resulting in
\begin{align}
    \rho_G^0(-2) &= \frac{1}{2^{N}} \prod_{\bhexagon} \frac{1 + \prod_{\ell_R \in \bhexagon} E_{R,\ell_R}^X \otimes Z_{M,\ell_R}(-3)}{8} \prod_{\ell_G} \frac{1+E_{G,\ell_G}^Y\otimes Z_{M,\ell_G}(-2)}{2} \nonumber\\
    &= \frac{1}{2^{N+N_M}} \sum_{f_B^X} f_B^X \otimes Z_{M,f_B^X}(-3) \cdot \prod_{\ell_G} \frac{1+E_{G,\ell_G}^Y\otimes Z_{M,\ell_G}(-2)}{2}
\end{align}
where $N_M$ is the total number of ancilla qubits introduced in one round of measurement, $\ell_R \in \bhexagon$ runs over three red edges in the blue hexagon.
The ancilla operator $Z_{M,f_B^X}(-3) \equiv \otimes_{\ell_R \in f_B^X } Z_{M,\ell_R}(-3)$ with $\ell_R \in f_B^X$ running over all the red edges within the support of the membrane operator $f_B^X$.
In the stabilizer expansion of $\rho_R^0(-3)$, only the operators that commute with all $E_{G,\ell_G}^Y$ survive after the measurements of $E_G^Y$.
Such operators $f_B^X$ are the product of the six-qubit vertex operators $V_B^X$.
Note that $f_B$ and $f_B^X$ differ by a product of $E_{G,\ell_G}^Y$ check operators.

At $t=-1$, we measure the check operators $E_{B,\ell_B}^Z$ in the state $\rho_G^0(-2)$, leading to
\begin{align}
    \rho_B^0(-1) = \frac{1}{2^{N+2N_M}} \sum_{f_B,f_R^Y} f_B f_R^Y \otimes Z_{M,f_B}(-3) \otimes Z_{M,f_B}(-2) \otimes Z_{M,f_R^Y}(-2) 
    \cdot \prod_{\ell_B} \frac{1+E_{B,\ell_B}^Z\otimes Z_{M,\ell_B}(-1)}{2}. \label{eq:rho_B0}
 \end{align}
Only the stabilizers that commute with all $E_{B,\ell_B}^Z$ will survive.
The membrane operator $f_B^X$ in general anti-commutes with $E_{B,\ell_B}^Z$ operators, while $f_B$ given by the product of $f_B^X$ and $E_{G,\ell_G}^Y$ commutes with $E_{B,\ell_B}^Z$.
In addition, the membrane operator $f_R^Y$, which is a product of six-qubit vertex operators $V_R^Y$ (a product of $E_{G,\ell_G}^Y$), also commutes with all $E_{B,\ell_B}^Z$.
Such an operator $f_R^Y$ differs from $f_R$ by a product of $E_{B,\ell_B}^Z$ check operators.
Lastly, we perform another round of $E_{R,\ell_R}^X$ measurement to obtain the maximally-mixed state in the logical space at time step $t = 0$.
The membrane operators $f_B$ are generated by the stabilizers $V_B^Z$ and commute with all check operators.
The products of $f_R^Y$ and $E_{B,\ell_B}^Z$ form the operators $f_R$ that commute with all $E_{R,\ell_R}^X$.
The operator $f_G^Z$ generated by $V_G^Z$ also survive projection onto the eigenbases of $E_{R,\ell_R}^X$, and are equivalent to $f_G$ up to the $E_{R,\ell_R}^X$ operators.
This leads to the state at time $t=0$
\begin{equation}
    \rho_R^0(0) = \frac{1}{2^{N+3N_M}} \sum_{f_R,f_G^Z,f_B} f_R f_G^Z f_B \otimes \calM(-1) \cdot \prod_{\ell_R} \frac{1+E_{R,\ell_R}^X\otimes Z_{M,\ell_R}(0)}{2},\label{eq:warm-up}
\end{equation}
where the ancilla qubits are included in
\begin{align}
    \calM(-1) = Z_{M,f_B}(-3) \otimes Z_{M,f_B}(-2) 
    \otimes Z_{M,f_R}(-2) \otimes Z_{M,f_R}(-1) \otimes Z_{M,f_G^Z}(-1).
\end{align}

\subsection{Decohered state in the stabilizer expansion} \label{Appendix:corrupted_state_derivation}
In this section, we derive the effects of the decoherence channel on the states at each time step of the measurement evolution.
We first apply the decoherence channel in \eqnref{eq:RGB_channels} to the initial state in \eqnref{eq:warm-up}, giving $\rho_R^0(0) \to \rho_R(0) \equiv \calN[\rho_R^0(0)]$.
These results can be easily generalized to any state during the evolution.
In the remainder of this section, we use $f_b$ to denote both the membrane operators and their configurations.

We first consider the errors $E_R^Y$, $E_R^Z$, $E_B^X$ and $E_G^X$ that commute with the $E_R^X$ check operators.
The $b$-color error types only anticommute with the membrane operator $f_b$ of the same color.
For $E_{R,\ell_R}^Y$ and $E_{R,\ell_R}^Z$ occurring on a red edge $\ell_R$, the error channel is given as
\begin{equation}
    \calN_{R,\ell_R}^Y \circ \calN_{R,\ell_R}^Z[\cdot] = (1-p)^2 [\cdot] + p(1-p)E_{R,\ell_R}^Y [\cdot] E_{R,\ell_R}^Y 
    + p(1-p) E_{R,\ell_R}^Z [\cdot] E_{R,\ell_R}^Z + p^2 E_{R,\ell_R}^X [\cdot] E_{R,\ell_R}^X.
\end{equation}
These error channels act on the red membranes $f_R$ as
\begin{equation}
    \calN_{R,\ell_R}^Y \circ \calN_{R,\ell_R}^Z[f_R] = \bigg( 1-2p(1-p) \bigg) f_R + 2p(1-p) E_{R,\ell_R}^Y f_R E_{R,\ell_R}^Y.
\end{equation}
Here, $f_R$ commutes with $E_{R,\ell_R}^X$, and $E_{R,\ell_R}^Y$ and $E_{R,\ell_R}^Z$ errors are equivalent in round R.
For a single red edge $\ell_R$, the effect of the channel on $f_R$ is
\begin{align*}
    \calN_{R,\ell_R}^Y \circ \calN_{R,\ell_R}^Z[f_R] = 
    \begin{cases}
        (1-2p)^2 f_R & \ell_R \in \partial f_R \\
        f_R & \ell_R \notin \partial f_R
    \end{cases}
\end{align*}
Hence, the channel $\calN_R^Y \circ \calN_R^Z \equiv \circ_{\ell_R} \calN_{R,\ell_R}^Y \circ \calN_{R,\ell_R}^Z$  assigns a real positive weight on $f_R$ according to the size of its boundary $\partial f_R$
\begin{equation}
    \calN_R^Y \circ \calN_R^Z[f_R] = e^{-2\mu |\partial f_R|} f_R
\end{equation}
with $\mu=-\log(1-2p)$ and $|\partial f_R|$ represents the total length of $\partial f_R$.

The $E_B^X$ and $E_G^X$ errors only anticommute with the blue membranes $f_B$ and green membranes $f_G^Z$, respectively.
On a single green edge $\ell_G$ and a single blue edge $\ell_B$, the error channels are given as 
\begin{align}
    \calN_{G,\ell_G}^X[\cdot] &= (1-p) [\cdot] + p E_{G,\ell_G}^X [\cdot] E_{G,\ell_G}^X \nonumber \\
    \calN_{B,\ell_B}^X[\cdot] &= (1-p) [\cdot] + p E_{B,\ell_B}^X [\cdot] E_{B,\ell_B}^X.
\end{align}
The single-edge channels assign a real positive weight to $f_B$ and $f_G^Z$, giving
\begin{align*}
    \calN_{G,\ell_G}^X [f_G^Z] = 
    \begin{cases}
        (1-2p) f_G^Z & \ell_G \in \partial f_G^Z \\
        f_G^Z & \ell_G \notin \partial f_G^Z
    \end{cases}
\end{align*}
and
\begin{align*}
    \calN_{B,\ell_B}^X [f_B] = 
    \begin{cases}
        (1-2p) f_B & \ell_B \in \partial f_B \\
        f_B & \ell_B \notin \partial f_B
    \end{cases}
\end{align*}
The channels on all green and blue edges acting on the $f_G^Z$ and $f_B$ operators are
\begin{align}
    \calN_G^X[f_G^Z] &= e^{-\mu |\partial f_G^Z|} f_G^Z \nonumber\\
    \calN_B^X[f_B] &= e^{-\mu |\partial f_B|} f_B.
\end{align}

Next, we consider the errors $E_G^Z$ and $E_B^Y$ that anticommute with the $E_R^X$ check operators.
We measure the corrupted state $\rho_R(0)$ in the basis of the $E_G^Y$ check operators at time step $t=1$.
Since the $E_G^Z$ and $E_B^Y$ errors always commute with the $E_G^Y$ check operators, we only need to consider which combinations of the $E_R^X$ check operators will survive the measurement.
The newly generated membrane operators are $f_B^X$ and $f_G$, which are the product of $f_G^Z$ and $E_R^X$ check operators.
The operators $E_G^Z$ and $E_B^Y$ only anticommute with the $f_G$ and $f_B f_B^X$ operators, respectively.
Hence, we arrive at
\begin{align}
    \calN_G^Z[f_G] &= e^{-\mu|\partial f_G|} f_G \nonumber\\
    \calN_B^Y[f_B f_B^X] &= e^{-\mu |\partial (f_B f_B^X)|} f_B f_B^X.
\end{align}

The state after applying the decoherence channel and $E_G^Y$ measurements at $t=1$ is
\begin{equation}
    \rho_G^0(1) = \frac{1}{2^{N+4N_M}} \sum_{f_R,f_G,f_B,f_B^X} w(0) f_R f_G f_B f_B^X \otimes \calM(-1) \cdot \prod_{\ell_G} \frac{1+E_{G,\ell_G}^Y \otimes Z_{M,\ell_G}(1)}{2} \label{eq:rho_R_start}
\end{equation}
where the weight is given by
\begin{equation}
    w(0) = e^{-2\mu |g_R| -2\mu|g_G| -\mu|g_B| -\mu|g_B+g_B^X|}.
\end{equation}
Here and in what follows, we introduce 
\begin{align}
g_b = \partial f_b,
\end{align}
which is a collection of binary variables representing the boundary of the membrane operator $f_b$.
Such binary variables live on the edges of the original triangular lattice, i.e. $g_{b,\ell} = 1$ if $\ell \in \partial f_b$. 
Note that the boundary of $f_Bf_B^X$ is given by the elementwise binary sum of $g_B$ and $g_B^X$.

\subsection{Measurement evolution details} \label{Appendix: evolution details}
In this section, we formulate the states at each step of the measurement evolution and derive the states $\rho_1$ and $\rho_2$ in the probe and $\rho_{\QM}$ and $\rho_{\QMR}$ in the coherent information. 

\subsubsection{$\rho_1$ and $\rho_2$}
We write the states at each time step of the evolution in \secref{Section: evolution}. 
As we go through the process, we can explicitly see how $m$-type logical operators are mapped to $e$-type logical operators after odd number of periods.
To find $\rho_1$, we start in the state in \eqnref{eq:warm-up} and follow the sequence listed in \eqnref{eq:rho12_period}. 
First, we initialize the state in the eigenbasis of the logical operator $L_B^X$, arriving at
\begin{equation}
    \rho_{R,1}^0(0) = \rho_R^0(0) \frac{1+L_B^X \otimes Z_L}{2}.
\end{equation}

\begin{figure*}[t]
\centering
\includegraphics[width=0.25\linewidth]{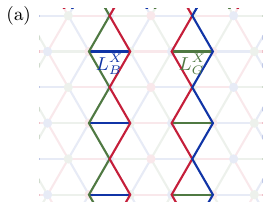}
\includegraphics[width=0.25\linewidth]{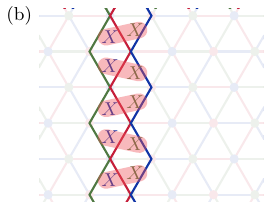}
    \caption{Logical operators enclosed by links on the original lattice (left panel) and an illustration of how $L_B^X$ evolves to $L_G^X$ by forming products with the $E_R^X$ check operators (right panel). $L_B^X$ can be seen as a homologically non-trivial strip enclosed by red and green links, while $L_G^X$ is enclosed by red and blue links. Through measurement evolution, $E_R^X$ checks highlighted in red act on $L_B^X$, transforming it into $L_G^X$.}
    \label{fig:logical_honeycomb_evolution}
\end{figure*}

We then apply the decoherence channel to $\rho_{R,1}^0(0) \to \rho_{R,1}(0) \equiv \calN[\rho_{R,1}^0(0)]$. 
Out of the errors that commute with the $E_R^X$ check operators, the errors $E_R^Y$ and $E_R^Z$ can anticommute with $L_B^X$.
After measuring the $E_G^Y$ check operators at $t=1$, the logical operator $L_B^X$ transforms into $L_G^X$ by forming products with the $E_R^X$ operators. 
The errors $E_B^Y$ can anticommute with $L_G^X$.
The state then evolves to 
\begin{align}
    \rho_{G,1}^0(1) =& \frac{1}{2^{N+4N_M+1}} \sum_{f_R,f_G,f_B,f_B^X} w(0) f_R f_G f_B f_B^X \bigg(1 + \lambda(0) L_G^X \otimes Z_L \otimes Z_{M,L_{B,R}^X}(0) \bigg) \otimes \calM(0) \nonumber \\
    &\cdot \prod_{\ell_G}\frac{1+E_{G,\ell_G}^Y\otimes Z_{M,\ell_G}(1)}{2}
\end{align}
where $\calM(0) = \calM(-1) \otimes Z_{M,f_G}(0) \otimes Z_{M,f_B^X}(0)$, and the weights are given by
\begin{align}
    w(0) &= e^{-2\mu |g_R| - 2\mu |g_G| - \mu|g_B|-\mu|g_B + g_B^X|} \nonumber \\
    \lambda(0) &= e^{-2\mu | (\partial L_B^X)_R + g_R | + 2\mu |g_R|} e^{-\mu |(\partial L_G^X)_B + g_B + g_B^X| + \mu |g_B + g_B^X |}.
\end{align}
The logical operator $L_B^X$ ($L_G^X$) is a strip that is enclosed by red and green (red and blue) edges as shown in \figref{fig:logical_honeycomb_evolution}.
Here, $(\partial L_B^X)_R$ ($(\partial L_G^X)_B$) is a set of binary variables that represents the set of red (blue) edges on the boundary of $L_B^X$ ($L_G^X$).
The ancilla operator $Z_{M,L_{B,R}^X}(0) \equiv \otimes_{\ell_R \in L_{B,R}^X} Z_{M,\ell_R}(0)$ runs over all red edges $\ell_R$ that belong to the set of red edges that encloses $L_B^X$.

Next, we apply the error channel $\rho_{G,1}^0(1) \to \calN[\rho_{G,1}^0(1)]$ again.
The errors that commute with the $E_G^Y$ check operators are $E_G^X$, $E_G^Z$, $E_B^Y$ and $E_R^Y$.
Among these errors, only $E_B^Y$ and $E_R^Y$ can anticommute with the logical operator $L_G^X$.
To find the weights from the channels $\calN_B^X$ and $\calN_R^Z$, we need to evolve to round B.
Since $L_G^X$ anticommutes with some $E_B^Z$ check operators, it transforms into $L_G^Z$ by forming products with the $E_G^Y$ operators.
The $E_B^X$ errors can anticommute with $L_G^Z$.
The $E_G^Y$ operators form red membrane operators $f_R^Y$ and the blue operators $f_B^X$ transform into $f_B$.
We label the membranes operators that evolve from $f_B^X$ as $f_{B,2}$ and rename the original $f_B$ as $f_{B,1}$.
The expression for round B is
\begin{align}
    \rho_{B,1}^0(2) =& \frac{1}{2^{N+5N_M+1}} \sum_{f_R,f_R^Y,f_G,f_{B,1},f_{B,2}} \Bigg[ w(0) w(1) f_R f_R^Y f_G f_{B,1} f_{B,2} \otimes \calM(1)
    \nonumber \\
    & \bigg(1 + \lambda(0) \lambda(1) L_G^Z \otimes Z_L \otimes Z_{M,L_{B,R}^X}(0) \otimes Z_{M,L_G^X}(1) \bigg) \Bigg] \cdot \prod_{\ell_B} \frac{1+E_{B,\ell_B}^Z\otimes Z_{M,\ell_B}(2)}{2}
\end{align}
where $\calM(1) = \calM(-1) \otimes Z_{M,f_G}(0) \otimes Z_{M,f_{B,2}}(0) \otimes Z_{M,f_{B,2}}(1) \otimes Z_{M,f_R^Y}(1)$, and the weights are given by
\begin{align}
    w(1) &= e^{ -\mu |g_R| -\mu |g_R + g_R^Y| -2\mu |g_G| -2\mu |g_{B,1} + g_{B,2}|} \nonumber\\
    \lambda(1) &= e^{-2\mu |(\partial L_G^Z)_B + g_{B,1} + g_{B,2}| + 2\mu |g_{B,1} + g_{B,2}|} e^{-\mu |(\partial L_G^X)_R + g_R| + \mu |g_R|}.
\end{align}
The ancilla operator $Z_{M,L_G^X}(1) \equiv \otimes_{\ell_G \in L_G^X} Z_{M,\ell_G}(1)$ runs over all the green edges within $L_G^X$.

Then, we let $\rho_{B,1}^0(2) \to \calN[\rho_{B,1}^0(2)]$ undergo the error channel and then evolve to round R.
The errors that commute with the $E_B^Z$ operators are $E_B^X$, $E_B^Y$, $E_R^Z$, and $E_G^Z$ and only $E_B^X$ and $E_B^Y$ can anticommute with $L_G^Z$.
After measuring the $E_R^X$ check operators, $L_G^Z$ transforms into $L_R^Z$, which anticommutes with the $E_G^X$ errors.
The newly generated membrane operators are $f_G^Z$ and $f_R$ which is the product of the $f_R^Y$ and $E_B^Z$ operators.
We label the red operators from $f_R^Y$ as $f_{R,2}$ and the original $f_R$ as $f_{R,1}$.
Since in round R $f_G$ and $f_G^Z$ are equivalent, we denote them as $f_{G,1}$ and $f_{G,2}$, respectively.
The resulting state is
\begin{align}
     \rho_{R,1}^0(3) =& \frac{1}{2^{N+6N_M+1}} \sum_{\{f_{b,\tau}\}} \Bigg[ w(0) w(1) w(2)  \prod_{b,\tau} f_{b,\tau} \otimes \calM(2) \nonumber \\
    & \bigg(1 + \lambda(0) \lambda(1) \lambda(2) L_R^Z  \otimes Z_L \otimes Z_{M,L_{B,R}^X}(0) \otimes Z_{M,L_G^X}(1) \otimes Z_{M,L_{G,B}^Z}(2) \bigg) \Bigg] \cdot \prod_{\ell_R} \frac{1+E_{R,\ell_R}^X\otimes Z_{M,\ell_R}(3)}{2} \label{eq:rho_R_3}
\end{align}
where $b \in \{R,G,B\}$, $\tau \in \{1,2\}$, and $\calM(2) = \calM(-1) \otimes Z_{M,f_{G,1}}(0) \otimes Z_{M,f_{B,2}}(0) \otimes Z_{M,f_{B,2}}(1) \otimes Z_{M,f_{R,2}}(1) \otimes Z_{M,f_{R,2}}(2) \otimes Z_{M,f_{G,2}}(2)$.
The weights from the error channel take the form
\begin{align}
    w(2) &= e^{-2\mu |g_{R,1} + g_{R,2}| -\mu |g_{G,1}| -\mu |g_{G,1} + g_{G,2}| -2\mu |g_{B,1} + g_{B,2}|}\nonumber\\
    \lambda(2) &= e^{-2\mu |(\partial L_G^Z)_B + g_{B,1} + g_{B,2}| + 2\mu |g_{B,1} + g_{B,2}|} e^{-\mu |(\partial L_R^Z)_G + g_{G,1} + g_{G,2}| + \mu |g_{G,1} + g_{G,2}|}.
\end{align}

Finally, apply the decoherence channel to $\rho_{R,1}^0(4) \to \rho_{R,1}(4)$ and measure the $E_G^Y$ check operators at $t=4$.
The operators generated after evolving to round G are $f_B^X$ and $f_G$ which are products of the $f_G^Z$ and $E_R^X$ operators.
We label the $f_B^X$ operators as $f_{B,3}$ since $f_B^X$ and $f_B$ are equivalent in round G.
Among the errors that commute with the $E_R^X$ operators, the $E_B^X$ and $E_G^X$ operators can anticommute with $L_R^Z$.

We arrive at the state
\begin{align}
    \rho_1 \equiv \rho_{G,1}^0(4) = & \frac{1}{2^{N+7N_M+1}} \sum_{\{f_{b,\tau}\}} \Bigg[ w(0) w(1) w(2) w(3) \prod_{b,\tau} f_{b,\tau} \otimes \calM(3) \bigg(1 + \lambda(0) \lambda(1) \lambda(2) \lambda(3) L_R^Y  \otimes Z_L  \nonumber \\
    &\otimes Z_{M,L_{B,R}^X}(0) \otimes Z_{M,L_G^X}(1) \otimes Z_{M,L_{G,B}^Z}(2) \otimes Z_{M,L_R^Z}(3) \bigg) \Bigg] \cdot \prod_{\ell_G} \frac{1+E_{G,\ell_G}^Y \otimes Z_{M,\ell_G}(4)}{2}
\end{align}
where the logical operator $L_R^Z$ transforms into $L_R^Y$, both of which are $e$-type logical operators.
The $E_G^Z$ errors can anticommute with $L_R^Y$.
We denote the the ancilla operators from the membrane operators as
\begin{align}
    \calM(3) =& \calM(-1) \otimes Z_{M,f_{G,1}}(0) \otimes Z_{M,f_{B,2}}(0) 
    \otimes Z_{M,f_{B,2}}(1) \otimes Z_{M,f_{R,2}}(1)
    \otimes Z_{M,f_{R,2}}(2) \otimes Z_{M,f_{G,2}}(2) \nonumber \\
    & \otimes Z_{M,f_{G,2}}(3) \otimes Z_{M,f_{B,3}}(3).
\end{align}
In addition to the membrane operators $f_{b,\tau}$ in \eqnref{eq:rho_R_3}, there is one more blue membrane operator $f_{B,3}$.
The weights gained from the decoherence channel are
\begin{align}
    w(3) &= e^{-2\mu |g_{R,1} + g_{R,2}| -2\mu |g_{G,1} + g_{G,2}| -\mu |g_{B,1} + g_{B,2}| -\mu |g_{B,1} + g_{B,2} + g_{B,3}|} \nonumber \\
    \lambda(3) &= e^{-2\mu |(\partial L_R^Y)_G + g_{G,1} + g_{G,2}| +2\mu |g_{G,1} + g_{G,2}|} e^{-\mu |(\partial L_R^Z)_B + g_{B,1} + g_{B,2}| + \mu |g_{B,1} + g_{B,2}|}
\end{align}

The derivation of the final state $\rho_2$ is much easier, since we do not need to consider the evolution of logical operators.
We again start in the state in \eqnref{eq:rho_R_start} and arrive at
    \begin{align}
        \rho_{G,2}^0(4) =& \frac{1}{2^{N+7N_M}} \sum_{\{f_{b,\tau}\}} \Bigg[ w(0) w(1) w(2) w(3) \prod_{b,\tau} f_{b,\tau} \otimes \calM(3) \Bigg]
        \cdot \prod_{\ell_G} \frac{1+E_{G,\ell_G}^Y \otimes Z_{M,\ell_G}(4)}{2} \label{eq:rho_G2_4}
    \end{align}
which has the same positive weights $w(t)$ as $\rho_1$. The indices of the membrane operators $f_{b,\tau}$ run over $b \in \{R,G,B\}$ and $\tau \in \{1,2\}$ and also $f_{B,3}$.

After projecting $\rho_{G,2}^0(4)$ onto the eigenbasis of $L_R^Y$, we reach the final state
\begin{align}
    \rho_2 =& \frac{1}{2^{N+7N_M+1}} \sum_{\{f_{b,\tau}\}} \Bigg[ w(0) w(1) w(2) w(3) \prod_{b,\tau} f_{b,\tau} \otimes \calM(3) \bigg(1 + L_R^Y \otimes Z_L \bigg) \Bigg] 
    \cdot \prod_{\ell_G} \frac{1+E_{G,\ell_G}^Y\otimes Z_{M,\ell_G}(4)}{2}. \label{eq:rho_2}
\end{align}

To simplify the derivation later on, we express the weight $w(\tau)$ and $\lambda(\tau)$ in terms of the cumulative membrane/loop variable $\prod_\tau f_{b,\tau}$ for $b \in {R,G,B}$. Namely, we introduce the redefinition
\begin{align}
    \prod_{\tau'=1} f_{b,\tau'} \mapsto f_{b,\tau}.\label{eq:redefined_variables}
\end{align}
Accordingly, $\sum_{\tau'=1}^\tau g_{b,\tau'} \mapsto g_{b,\tau}$ (e.g. $g_{R,1} + g_{R,2} \mapsto g_{R,2}$).
We note that this redefinition does not change the summation over $f_{b,\tau}$ as the redefined $f_{b,\tau}$ are also independent.
Here, we summarize the results in terms of these redefined variables:
\begin{equation}
\begin{aligned}
    w(0) &= e^{-2\mu |g_{R,1}| - 2\mu |g_{G,1}| - \mu|g_{B,1}| - \mu|g_{B,2}|} 
    \\
    w(1) &= e^{ -\mu |g_{R,1}| -\mu |g_{R,2} | -2\mu |g_G| -2\mu | g_{B,2}|} 
    \\
    w(2) &= e^{-2\mu | g_{R,2}| -\mu |g_{G,1}| -\mu |g_{G,2}| -2\mu | g_{B,2}|}
    \\
    w(3) &= e^{-2\mu |g_{R,2}| -2\mu |g_{G,2}| -\mu |g_{B,2}| -\mu | g_{B,3}|} 
    \\
    \lambda(0) &= e^{-2\mu | (\partial L_B^X)_R + g_{R,1} | + 2\mu |g_{R,1}|} e^{-\mu |(\partial L_G^X)_B + g_{B,2}| + \mu |g_{B,2} |}
    \\
    \lambda(1) &= e^{-2\mu |(\partial L_G^Z)_B + g_{B,2}| + 2\mu |g_{B,2}|} e^{-\mu |(\partial L_G^X)_R + g_{R,1}| + \mu |g_{R,1}|}
    \\
    \lambda(2) &= e^{-2\mu |(\partial L_G^Z)_B + g_{B,2}| + 2\mu |g_{B,2}|} e^{-\mu |(\partial L_R^Z)_G + g_{G,2}| + \mu | g_{G,2}|}
    \\
    \lambda(3) &= e^{-2\mu |(\partial L_R^Y)_G + g_{G,2}| +2\mu |g_{G,2}|} e^{-\mu |(\partial L_R^Z)_B + g_{B,2}| + \mu |g_{B,2}|}
\end{aligned}
\end{equation}

%
\subsubsection{$\rho_{\QM}$ and $\rho_{\QMR}$}
To derive $\rho_{\QM}$, we start in the state $\rho_R^0(0)$ in \eqnref{eq:warm-up}.
After undergoing the evolution in \eqnref{eq:rho12_period}, we find
\begin{align}
    \rho_{\QM} = \frac{1}{2^{N+7N_M}} \sum_{\{f_{b,\tau}\}} \Bigg[ w(0) w(1) w(2) w(3) \prod_{b,\tau} f_{b,\tau} \otimes \calM(3) \Bigg]
     \cdot \prod_{\ell_G} \frac{1+E_{G,\ell_G}^Y \otimes Z_{M,\ell_G}(4)}{2}. \label{eq:rho_QM}
\end{align}
The details of the evolution are the same as in \eqnref{eq:rho_G2_4}.

The initial state of $\rho_{\QMR}$ is
\begin{equation}
    \rho_{\QMR}^0(0) = \prod_{l = l_1,l_2} \frac{1+L_{R,l}^Y Z^R_l}{2} \frac{1+L_{B,l}^X X_l^R}{2} \rho_R^0(0).
\end{equation}
The logical operator $L_R^Y$ is the Pauli-Z operator in the logical space while $L_B^X$ is the Pauli-X operator.

The density matrix $\rho_{\QMR}$ undergoes the same evolution as $\rho_{\QM}$ and takes the form in round G at $t=4$
\begin{align}
    \rho_{\QMR} =& \frac{1}{2^{N+7N_M+2}} \sum_{\{f_{b,\tau}\}} \Bigg\{ 
    \sum_{d_{l_1}^X,d_{l_2}^X = 0,1} \Bigg[ \prod_{t = 0}^3\lambda_X(t) \prod_{l=l_{1,2}} \bigg(L_{R,l}^Z  \otimes \calM_l^X(3) \bigg)^{d_{l}^X} \Bigg] 
    \nonumber \\
    & \sum_{d_{l_1}^Z,d_{l_2}^Z = 0,1} \Bigg[ \prod_{t = 0}^3\lambda_Z(t)
    \prod_{l=l_{1,2}} \bigg(L_{G,l}^X \otimes \calM_l^Z(3) \bigg)^{d_l^Z} \Bigg]\prod_{f_{b,\tau}}f_{b,\tau} \otimes \calM(3) \Bigg\} 
     \prod_{\ell_G} \frac{1+E_{G,\ell_G}^Y\otimes Z_{M,\ell_G}(4)}{2},
\end{align}
where the binary variables $d_{l_{1,2}}^{X,Z} \in \{0,1\}$ indicate whether a Pauli-X or Pauli-Z logical operator acts on the state in either direction $l_{1,2}$ of the torus.
We define the ancilla operators
\begin{align}
    \calM_l^X(3) &= Z_{M,L_{B,R,l}^X}(0) \otimes Z_{M,L_{G,l}^X}(1) \otimes Z_{M,L_{G,B,l}^Z}(2) \otimes Z_{M,L_{R,l}^Z}(3)
    \nonumber \\
    \calM_l^Z(3) &= Z_{M,L_{R,G,l}^Y}(1) \otimes Z_{M,L_{B,l}^Y}(2) \otimes Z_{M,L_{B,R,l}^X}(3)
    \nonumber.
\end{align}
The logical Pauli-X(Z) operators are given by $L_B^X$($L_R^Y$), whose evolution sequence is shown in \eqnref{eq:coherent_info_logical_sequence}. 
The weights $\lambda_Z$ in terms of the redefined variables in Eq.~\eqref{eq:redefined_variables} take the form
\begin{align}
    \lambda_Z(0) &= e^{-2\mu \Big| g_{G,1} + \sum_l [(\partial L_{R,l}^Y)_G]^{d_l^Z} \Big| -\mu \Big| g_{B,1} + \sum_l [(\partial L_{R,l}^Y)_B]^{d_l^Z} \Big|}
    \nonumber \\
    \lambda_Z(1) &= e^{-\mu \Big| g_{R,2} + \sum_l[(\partial L_{B,l}^Y)_R]^{d_l^Z} \Big| -2\mu \Big| g_{G,1} + \sum_l[(\partial L_{R,l}^Y)_G]^{d_l^Z} \Big|}
    \nonumber \\
    \lambda_Z(2) &= e^{-2\mu \Big| g_{R,2} + \sum_l [(\partial L_{B,l}^X)_R]^{d_l^Z} \Big| -\mu \Big| g_{G,1} + \sum_l [(\partial L_{B,l}^Y)_G]^{d_l^Z} \Big|}
    \nonumber \\
    \lambda_Z(3) &= e^{-2\mu \Big| g_{R,2} + \sum_l [(\partial L_{B,l}^X)_R]^{d_l^Z} \Big| -\mu \Big| g_{B,3} + \sum_l [(\partial L_{G,l}^X)_B]^{d_l^Z} \Big| }
\end{align}
and the derivation can be found in the evolution details of $\rho_1$.
At each time step, the errors that anticommute with the Pauli-X logical operators are those that commute with the Pauli-Z logical operators.
The $\lambda_X$ weights at times $t \in [0,3]$ are
\begin{align}
    \lambda_X(0) &= e^{-2\mu \Big| g_{R,1} + \sum_l [(\partial L_{B,l}^X)_R]^{d_l^X} \Big| -\mu \Big|g_{B,2} + \sum_l [(\partial L_{G,l}^X)_B]^{d_l^X} \Big|}
    \nonumber \\
    \lambda_X(1) &= e^{-\mu \Big| g_{R,1} + \sum_l [(\partial L_{G,l}^X)_R]^{d_l^X} \Big| -2\mu \Big| g_{B,2} + \sum_l [(\partial L_{G,l}^Z)_B]^{d_l^X} \Big|}
    \nonumber \\
    \lambda_X(2) &= e^{-\mu \Big| g_{G,2} + \sum_l [(\partial L_{R,l}^Z)_G]^{d_l^X} \Big| -2\mu \Big|g_{B,2} + \sum_l [(\partial L_{G,l}^Z)_B]^{d_l^X} \Big|}
    \nonumber \\
    \lambda_X(3) &= e^{-2\mu \Big| g_{G,2} + \sum_l [(\partial L_{R,l}^Y)_G]^{d_l^X} \Big| -\mu \Big| g_{B,2} + \sum_l [(\partial L_{R,l}^Z)_B]^{d_l^X} \Big|}.
\end{align}

\subsection{Mapping to ($n-1$)-flavor Ising model} \label{Appendix: Ising mapping}
In this section, we first calculate the $n$-th spectral moment of $\rho_{\QM}$ and write it as a product of the partition functions of the membrane configuration $f_{b,\tau}$ with the indices $b$ and $\tau$ running over the seven membrane operators in \eqnref{eq:rho_QM}.
Following \cite{Fan2024,hauser2024}, we show that these partition functions can be mapped to those of the ($n-1$)-flavor Ising model.

We start by finding the $n$-th spectral moment of $\rho_{\QM}$ in \eqnref{eq:rho_QM}
\begin{align}
    \tr \rho_{\QM}^n =& \frac{1}{2^{n(N+7N_M)}} \sum_{\{f_{b,\tau}^{(r)}\}}  \prod_{r=1}^n \Bigg( w^{(r)}(0) w^{(r)}(1) w^{(r)}(2) w^{(r)}(3) \Bigg) 
    \nonumber \\
    &\tr\Bigg\{ \prod_{r=1}^n \Bigg[ \prod_{b,\tau} f_{b,\tau}^{(r)} \otimes \calM^{(r)}(3)  
    \cdot \prod_{\ell_G} \frac{1+E_{G,\ell_G}^Y \otimes Z_{M,\ell_G}(4)}{2} \Bigg] \Bigg{\}} 
\end{align}
where $r=1,\cdots,n$ labels the $r$-th replica.
The non-vanishing trace terms have to be products proportional to identity because Pauli matrices have zero trace.

%
The membrane operators $f_{b,\tau}^{(r)}$ form seven independent constraints
\begin{equation}
    f_{b,\tau}^{(n)} = \prod_{r=1}^{n-1} f_{b,\tau}^{(r)}. \label{eq:loop_constraints}
\end{equation}
The operators $f_{b,\tau}$ are independent of each other because every $f_{b,\tau}$ is coupled with a tensor product of Pauli-$Z$ acting on the ancilla space.
Thus, the $n$-th moment is
\begin{align}
    \tr\rho_{\QM}^n =& \frac{1}{2^{(n-1)(N+7N_M)}} \label{eq:nth_moment_rho_QM}
    \sum_{\{f_{b,\tau}^{(r)}\}} \prod_{r=1}^n \Bigg( w^{(r)}(0) w^{(r)}(1) w^{(r)}(2) w^{(r)}(3) \Bigg).
\end{align}

We now take a closer look at the positive weights in terms of the constraints in \eqnref{eq:loop_constraints}.
The $n$-th moment factorizes into seven partition functions
\begin{equation}
    \tr \rho_{\QM}^n = \frac{1}{2^{(n-1)(N+7N_M)}} \calZ_{R,1} \calZ_{R,2} \calZ_{G,1} \calZ_{G,2} \calZ_{B,1} \calZ_{B,2} \calZ_{B,3},
\end{equation}
where
\begin{equation}
\begin{aligned}
     \calZ_{R,1} &= \sum_{ \{f_{R,1}^{(r)}\} } e^{-3\mu \big( \sum_{r=1}^{n-1} \big|g_{R,1}^{(r)}\big| + \big| \sum_{r=1}^{n-1} g_{R,1}^{(r)} \big| \big)} 
     \\
     \calZ_{G,1} &= \sum_{ \{f_{G,1}^{(r)}\} } e^{-5\mu \big( \sum_{r=1}^{n-1} \big|g_{G,1}^{(r)}\big| + \big| \sum_{r=1}^{n-1} g_{G,1}^{(r)} \big| \big)} 
     \\
     \calZ_{B,1} &= \sum_{ \{f_{B,1}^{(r)}\} } e^{-\mu \big( \sum_{r=1}^{n-1} \big|g_{B,1}^{(r)}\big| + \big| \sum_{r=1}^{n-1} g_{B,1}^{(r)} \big| \big)} 
     \\
    \calZ_{R,2} &= \sum_{ \{f_{R,2}^{(r)}\} } e^{-5\mu \big( \sum_{r=1}^{n-1} \big|g_{R,2}^{(r)}\big| + \big| \sum_{r=1}^{n-1} g_{R,2}^{(r)} \big| \big)} 
    \\
    \calZ_{G,2} &= \sum_{ \{f_{G,2}^{(r)}\} } e^{-3\mu \big( \sum_{r=1}^{n-1} \big|g_{G,2}^{(r)}\big| + \big| \sum_{r=1}^{n-1} g_{G,2}^{(r)} \big| \big)} 
    \\
    \calZ_{B,2} &= \sum_{ \{f_{B,2}^{(r)}\} } e^{-6\mu \big( \sum_{r=1}^{n-1} \big|g_{B,2}^{(r)}\big| + \big| \sum_{r=1}^{n-1} g_{B,2}^{(r)} \big| \big)} 
    \\
    \calZ_{B,3} &= \sum_{\{ f_{B,3}^{(r)} \} } e^{-\mu \big( \sum_{r=1}^{n-1} \big|g_{B,3}^{(r)}\big| + \big| \sum_{r=1}^{n-1} g_{B,3}^{(r)} \big| \big)} 
\end{aligned}\label{eq:Z1}
\end{equation}
where we recall that the loop configuration $g_{b,\tau}^{(r)} = \partial f_{b,\tau}^{(r)}$.

%
The prefactors in the partition functions $\calZ_{b,\tau}$ should always be $\mu_{b,\tau}=6\mu$ if we allow the evolution to continue running.
The reason why the prefactors in \eqnref{eq:Z1} are different from $6\mu$ is that in the warm-up period, no errors occurred and we only kept track of the errors occurring until after round R at time $t=3$.
For at least one partition function to accumulate a prefactor of $6\mu$, the decoherence channel needs to be applied at a minimum of four consecutive time steps. 
If the decoherence channel is applied less than four times, the partition functions are mapped to ($n-1$)-flavor Ising models with a higher transition threshold.
Note that $\mu_{b,\tau}$ is proportional to the number of times the $b$-color error is applied as shown in \figref{fig:simple_error_detection}.

Each of the partition function can be written as
\begin{align}
    \calZ_{b,\tau} = \sum_{ \{f_{b,\tau}^{(r)}\} } e^{-H_{b,\tau}}
\end{align}
where the positive weights can be expressed as Hamiltonians $H_{b,\tau}$.

\begin{figure}
\centering
\includegraphics[width=0.25\linewidth]{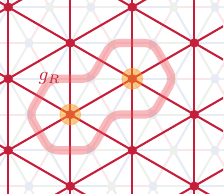}
\caption{The ($n-1$)-flavor Ising model on the triangular lattice. The model involves $n-1$ flavors of the Ising spins defined on red vertices. The domain wall of the Ising spins represents the loop configuration $g_R$ in the expansion of the density matrix.}
\label{fig:Ising_spins}
\end{figure}

We now write the partition functions of membrane configurations as those of the ($n-1$)-flavor Ising model.
We introduce the Ising spins $\sigma_{b,\tau,i}^{(r)} \in \{-1,+1\}$ on the $b$-colored vertices of the original lattice as shown in Fig.~\ref{fig:Ising_spins}.
The index $i$ indicates the location of the spin.
The boundary of $b$-colored membrane operator $g_{b,\tau}^{(r)}$ are identified as the domain walls of the Ising spins
\begin{equation}
    g_{b,\tau,l}^{(r)} = \frac{1-\sigma_{b,\tau,i}^{(r)} \sigma_{b,\tau,j}^{(r)}}{2} 
\end{equation}
where the link connecting the Ising spins at locations $i$ and $j$ is dual to $l$.
The binary function $g_{b,\tau,l}^{(r)} \in \{0,1\}$ is defined as 
\begin{equation}
    g_{b,\tau,l}^{(r)} =
    \begin{cases}
         1 & l \in g_{b,\tau}^{(r)} \\
         0 & l \notin g_{b,\tau}^{(r)} \\
    \end{cases}
\end{equation}
with the total length of a membrane operator boundary represented by $|g_{b,\tau}^{(r)}| = \sum_l g_{b,\tau,l}^{(r)}$.

Then, the effective Hamiltonian (for $g_{B,2}^{(r)}$) in terms of the Ising spins is
\begin{align}
    H_{b,\tau} = -3\mu \sum_{\langle i,j \rangle} \bigg( & \sum_{r=1}^{n-1} \sigma_{b,\tau,i}^{(r)} \sigma_{b,\tau,j}^{(r)} 
     + \prod_{r=1}^{n-1} \sigma_{b,\tau,i}^{(r)} \sigma_{b,\tau,j}^{(r)} \bigg).
\end{align}

\subsection{Statistical mechanics mapping for diagnostics} \label{Appendix: stat-mech diagnostics}
In this section, we show that the $e$-$m$ automorphism probe and the Rényi coherent information are associated with the excessive free energy of inserting defects along a homologically non-trivial loop around one direction of the torus. 

\subsubsection{Probe of $e$-$m$ automorphism}
The Rényi relative entropy is
\begin{equation}
    D_{em}^{(n)}(\rho_2 || \rho_1) = \frac{1}{1-n} \log\frac{\tr \rho_2 \rho_1^{n-1}}{\tr \rho_2^n}
\end{equation}

We start by finding the $n$-th spectral moment of $\rho_2$
    \begin{align}
        \tr \rho_2^n =& \frac{1}{2^{n(N+7N_M+1)}} \sum_{\{f_{b,\tau}^{(r)}\}}  \prod_{r=1}^n \Bigg( w^{(r)}(0) w^{(r)}(1) w^{(r)}(2) w^{(r)}(3) \Bigg) \tr\Bigg\{ \prod_{r=1}^n \Bigg[ \prod_{b,\tau} f_{b,\tau}^{(r)} \otimes \calM^{(r)}(3) \bigg(1 + L_R^Y \otimes Z_L \bigg) \Bigg] 
        \nonumber \\
        &  \cdot \prod_{\ell_G} \frac{1+E_{G,\ell_G}^Y \otimes Z_{M,\ell_G}(4)}{2} \Bigg] \Bigg{\}}.
    \end{align}
The logical operator $L_R^{Y,(r)}$ cannot be annihilated by the membrane operators and only terms with even copies of $L_R^{Y,(r)}$ remain. 
This contributes a factor of $2^{n-1}$ because we sum up the different ways of choosing even number copies of the logical operator. 
The trace of the identity matrix of the system, the measurement ancilla qubits from $t=-3$ to $t=3$ and the logical ancilla qubit is $2^{N+7N_M+1}$.
Thus, the $n$-th moment is
\begin{align}
    \tr\rho_2^n =& \frac{1}{2^{(n-1)(N+7N_M)}} \label{eq:nth_moment_rho2}
    \sum_{\{f_{b,\tau}^{(r)}\}} \prod_{r=1}^n w^{(r)}(0) w^{(r)}(1) w^{(r)}(2) w^{(r)}(3).
\end{align}

We initialize in the $+1$-eigenstate ($\ket{0}_L$) of the logical operator and the $+1$-eigenstate of the logical ancilla in the Z-basis.
After multiple rounds of measurements, the logical operator couples to Pauli-Z operators of the measurement ancilla qubits and we are no longer in the $+1$-logical eigenstate.
To see how this happens, we can start in the $+1$-eigenstate of $L_B^X$
\begin{align}
    \rho_+(t=0) =& \frac{1}{2^{N+3N_M}} \sum_{f_R,f_G^Z,f_B} f_R f_G^Z f_B \otimes \calM(-1) 
    \frac{1+L_B^X}{2} \otimes \frac{1+Z_L}{2} \cdot \prod_{\ell_R} \frac{1+E_{R,\ell_R}^X \otimes Z_{M,\ell_R}(0)}{2}
\end{align}
at $t=0$.
We let $\rho_+^0$ undergo GBRG measurements without local errors.
At $t=4$, the state becomes
\begin{align}
    \rho_+(t=4) = \frac{1}{2^{N+7N_M}} \sum_{\{f_{b,\tau}\}} \Bigg[& \prod_{b,\tau} f_{b,\tau} \bigg( \frac{1 + L_R^Y \otimes Z_{M,L_{B,R}^X}(0) \otimes Z_{M,L_G^X}(1) \otimes Z_{M,L_{G,B}^Z}(2) \otimes Z_{M,L_R^Z}(3)}{2} \otimes \frac{1+Z_L}{2} \bigg) 
    \nonumber \\
    &\otimes \calM(3) \Bigg] \cdot \prod_{\ell_G} \frac{1+E_{G,\ell_G}^Y\otimes Z_{M,\ell_G}(4)}{2}, \label{eq:rho_L4}
\end{align}
which is no longer the $+1$-eigenstate of $L_R^Y$ but this is not reflected in the logical ancilla qubit.
Since we are evolving the logical operator through measurements in $\rho_1$ but not in $\rho_2$, we will need to do a basis rotation in either of the states. 
The unitary operator for this rotation is 
\begin{equation}
    C = \frac{1+Z_C}{2} \otimes I_L + \frac{1-Z_C}{2} \otimes L_G^X  \label{eq:rot_meas}
\end{equation}
where $Z_C = Z_{M,L_{B,R}^X}(0) \otimes Z_{M,L_G^X}(1) \otimes Z_{M,L_{G,B}^Z}(2) \otimes Z_{M,L_R^Z}(3)$ and $I_L$ is the identity operator on the logical subspace.
The logical operator $L_G^X$ acts on a different direction of the torus than $L_R^Y$.
The action of $C$ is to decouple the logical subspace from the measurement ancillae so that the $\ket{0}_L, \ket{1}_L$ state of the logical ancilla qubit corresponds to the $+1$-, $-1$-eigenstate of $L_R^Y$.
Hence, we are actually interested in the Rényi relative entropy of $C\rho_1 C^{\dagger}$ and $\rho_2$.

We are now ready to find the numerator of the Rényi relative entropy
    \begin{align}
        \tr\bigg( \rho_2 (C\rho_1C^{\dagger})^{n-1} \bigg) =& \frac{2^{7N_M}}{2^{n(N+7N_M+1)}} \sum_{\{f_{b,\tau}^{(r)}\}}  \prod_{r=1}^n \Bigg( w^{(r)}(0) w^{(r)}(1) w^{(r)}(2) w^{(r)}(3) \Bigg)   \nonumber \\
        & \tr\Bigg{\{ } \underbrace{\bigg(1 + L_G^{X} \otimes Z_L \bigg)}_{\text{from } \rho_2} \prod_{r=2}^n \underbrace{\bigg(1 + \lambda^{(r)}(0) \lambda^{(r)}(1) \lambda^{(r)}(2) \lambda^{(r)}(3) L_G^{X} \otimes Z_L \bigg)}_{\text{from } C \rho_1 C^\dagger } \Bigg{ \} }
    \end{align}
where we have applied the constraints in \eqnref{eq:loop_constraints} and traced over measurement ancillae.
Now, only terms with even copies of logical operators will remain. 
Since there is a copy of $L_G^X$ from $\rho_2$, we can have any number of copies of $L_G^X$ contributed by $\rho_1$.
Thus, after accounting for the logical operators, we arrive at the expression
    \begin{align}
        \tr\bigg( \rho_2 (C\rho_1C^{\dagger})^{n-1} \bigg) = \frac{1}{2^{(n-1)(N+7N_M+1)}} \sum_{\{f_{b,\tau}^{(r)}\}} \Bigg[ &\prod_{r=1}^n \bigg( w^{(r)}(0) w^{(r)}(1) w^{(r)}(2) w^{(r)}(3) \bigg) 
        \nonumber \\
        &\prod_{r=2}^n \bigg(1 + \lambda^{(r)}(0) \lambda^{(r)}(1) \lambda^{(r)}(2) \lambda^{(r)}(3) \bigg) \Bigg]. \label{eq:numerator}
    \end{align}
We can write \eqnref{eq:numerator} in terms of partition functions
    \begin{equation}
    \tr\bigg( \rho_2 (C\rho_1C^{\dagger})^{n-1} \bigg) = \frac{1}{2^{(n-1)(N+7N_M+1)}} \calZ_{R,2} \calZ_{G,1} \calZ_{B,1} \calZ_{B,3} \sum_{\bf{d}} \calZ_{R,1}^{(\bf d)} \calZ_{G,2}^{(\bf d)} \calZ_{B,2}^{(\bf d)} 
    \end{equation}
where we defined $\bf{d}$ as a ($n-1$)-component binary vector with components $d_r \in \{0,1\}$ and $r=2,\cdots,n$.
The partition functions with logical operators are
\begin{align}
    \calZ_{R,1}^{(\bf{d})} &= \sum_{ \{f_{R,1}^{(r)}\} } e^{-3\mu \big( \sum_{r=2}^{n} \big|[(\partial L_G^{X,(r)})_R]^{d_r} + g_{R,1}^{(r)}\big| + \big| \sum_{r=2}^{n} g_{R,1}^{(r)} \big| \big)} 
    \nonumber\\
    \calZ_{G,2}^{(\bf{d})} &= \sum_{ \{f_{G,2}^{(r)}\} } e^{-3\mu \big( \sum_{r=2}^{n} \big|[(\partial L_R^{Z,(r)})_G]^{d_r} + g_{G,2}^{(r)}\big| + \big| \sum_{r=2}^{n} g_{G,2}^{(r)} \big| \big)}
    \\
    \calZ_{B,2}^{(\bf{d})} &= \sum_{ \{f_{B,2}^{(r)}\} } e^{-6\mu \big( \sum_{r=2}^{n} \big|[(\partial L_R^{Z,(r)})_B]^{d_r} + g_{B,2}^{(r)}\big| + \big| \sum_{r=2}^{n} g_{B,2}^{(r)} \big| \big)}
    \nonumber
\end{align}
and we have used the fact that logical operators $L_B^X$ and $L_G^X$ share the same red links after measurement evolution while $L_G^Z$ and $L_R^Z$ share the same blue links.

Hence, the Rényi relative entropy is 
\begin{equation}
    D_{em}^{(n)}(\rho_2 || C\rho_1 C^{\dagger}) = \frac{1}{1-n} \log{ \frac{\sum_{\bf{d}} \calZ_{R,1}^{(\bf{d})} \calZ_{G,2}^{(\bf{d})} \calZ_{B,2}^{(\bf{d})} }{ 2^{n-1} \calZ_{R,1} \calZ_{G,2} \calZ_{B,2} } }
    \label{eq:relative_entropy}
\end{equation}
and is associated with the free energy of inserting defects along homologically non-trivial loops
\begin{equation}
    \Delta F_{\bf{d}} \equiv -\log \frac{\calZ^{(\bf{d})}_{R,1} \calZ^{(\bf{d})}_{G,2} \calZ^{(\bf{d})}_{B,2} }{\calZ_{R,1} \calZ_{G,2} \calZ_{B,2} }.
\end{equation}
In the paramagnetic phase (i.e. $p < p_c$), the defect free energy is vanishing, $\Delta F_{\bf{d}} = 0$, leading to $D_{em}^{(n)} = 0$.
In the ferromagnetic phase (i.e. $p > p_c$), the defect free energy diverges, $F_{\bf{d}} \to \infty$ unless $\bf{d} = \bf{0}$, leading to $D_{em}^{(n)} = \log 2$.
We note that in the Toric code phase (e.g. in the modified measurement sequence explained at the end of Sec.~\ref{sec: setup_logicals}), logical operators in the expansion of $\rho_2$ and $C\rho_1 C^\dagger$ are not identical.
As a consequence, the summation in Eq.~\eqref{eq:relative_entropy} (inside the log) has an additional constraint on the $(n-1)$-component vector $\bf{d}$ that $\mathrm{sum}(\bf{d}) = 0 \mod 2$, and therefore only involves $2^{n-2}$ terms.
In the Toric code phase (a paramagnetic phase), this gives rise to $D_{em}^{(n)} = (\log 2)/(n-1)$.

\subsubsection{Coherent information}
The Rényi coherent information is 
\begin{equation}
    I_c^{(n)}(\mathrm{R}, \QM) = \frac{1}{n-1} \frac{\tr \rho^n_{\QMR}}{\tr \rho_{\QM}^n}
\end{equation}
and can be mapped to the excessive free energy of inserting defects along homologically non-trivial loops.

The $n$-th moment of $\rho_{\QMR}$ is
\begin{align}
    \tr \rho_{\QMR}^n = \frac{1}{2^{(n-1)(N+7N_M+2)}} \bigg[\sum_{\mathbf{d}^X_{l_1}, \mathbf{d}^X_{l_2}} \calZ_{R,1}^{(\mathbf{d}^X_{l_1}, \mathbf{d}^X_{l_2})} \calZ_{G,2}^{(\mathbf{d}^X_{l_1}, \mathbf{d}^X_{l_2})} \calZ_{B,2}^{(\mathbf{d}^X_{l_1}, \mathbf{d}^X_{l_2})} \bigg] \bigg[\sum_{\mathbf{d}^Z_{l_1}, \mathbf{d}^Z_{l_2}} \calZ_{R,2}^{(\mathbf{d}^Z_{l_1}, \mathbf{d}^Z_{l_2})} \calZ_{G,1}^{(\mathbf{d}^Z_{l_1}, \mathbf{d}^Z_{l_2})} \calZ_{B,1}^{(\mathbf{d}^Z_{l_1}, \mathbf{d}^Z_{l_2})} \calZ_{B,3}^{(\mathbf{d}^Z_{l_1}, \mathbf{d}^Z_{l_2})} \bigg]
\end{align}
where $\mathbf{d}_{l_{1,2}}^{X,Z}$ are binary ($n-1$)-component vectors whose $r$-th component is $d_{l_{1,2},r}^{X,Z}$.

The partition functions are 
\begin{align}
    \calZ_{R,1}^{(\mathbf{d}^X_{l_1}, \mathbf{d}^X_{l_2})} &= \sum_{ \{f_{R,1}^{(r)}\} } e^{-3\mu \Big\{ \sum_{r=1}^{n-1} \Big|\sum_l [(\partial L_{B,l}^{X,(r)})_R]^{d_{l,r}^X} + g_{R,1}^{(r)}\Big| + \Big| \sum_{r=1}^{n-1} \big( \sum_l [(\partial L_{B,l}^{X,(r)})_R]^{d_{l,r}^X} + g_{R,1}^{(r)} \big) \Big| \Big\} } 
    \nonumber \\
    \calZ_{R,2}^{(\mathbf{d}^Z_{l_1}, \mathbf{d}^Z_{l_2})} &= \sum_{ \{f_{R,2}^{(r)}\} } e^{-5\mu \Big\{ \sum_{r=1}^{n-1} \Big|\sum_l [(\partial L_{B,l}^{X,(r)})_R]^{d_{l,r}^Z} + g_{R,2}^{(r)} \Big| + \Big| \sum_{r=1}^{n-1} \big( \sum_l [(\partial L_{B,l}^{X,(r)})_R]^{d_{l,r}^Z} + g_{R,2}^{(r)} \big) \Big| \Big\} } 
    \nonumber \\
    \calZ_{G,1}^{(\mathbf{d}^Z_{l_1}, \mathbf{d}^Z_{l_2})} &= \sum_{ \{f_{G,1}^{(r)}\} } e^{-5\mu \Big\{ \sum_{r=1}^{n-1} \Big|\sum_l [(\partial L_{R,l}^{Y,(r)})_G]^{d_{l,r}^Z} + g_{G,1}^{(r)}\Big| + \Big| \sum_{r=1}^{n-1} \big( \sum_l  [(\partial L_{R,l}^{Y,(r)})_G]^{d_{l,r}^Z} + g_{G,1}^{(r)} \big) \Big| \Big\} }
    \\
    \calZ_{G,2}^{(\mathbf{d}^X_{l_1}, \mathbf{d}^X_{l_2})} &= \sum_{ \{f_{G,2}^{(r)}\} } e^{-3\mu \Big\{ \sum_{r=1}^{n-1} \Big|\sum_l [(\partial L_{R,l}^{Z,(r)})_G]^{d_{l,r}^X} + g_{G,2}^{(r)} \Big| + \Big| \sum_{r=1}^{n-1} \big( \sum_l [(\partial L_{R,l}^{Z,(r)})_G]^{d_{l,r}^X} + g_{G,2}^{(r)} \big) \Big| \Big\} } 
    \nonumber \\
    \calZ_{B,1}^{(\mathbf{d}^Z_{l_1}, \mathbf{d}^Z_{l_2})} &= \sum_{ \{f_{B,1}^{(r)}\} } e^{-\mu \Big\{ \sum_{r=1}^{n-1} \Big| \sum_l [(\partial L_{R,l}^{Y,(r)})_B]^{d_{l,r}^Z} + g_{B,1}^{(r)} \Big| + \Big| \sum_{r=1}^{n-1} \big( \sum_l [(\partial L_{R,l}^{Y,(r)})_B]^{d_{l,r}^Z} + g_{B,1}^{(r)} \big) \Big| \Big\} }
    \nonumber\\
     \calZ_{B,2}^{(\mathbf{d}^X_{l_1}, \mathbf{d}^X_{l_2})} &= \sum_{ \{f_{B,2}^{(r)}\} } e^{-6\mu \Big\{ \sum_{r=1}^{n-1} \Big|\sum_l [(\partial L_{R,l}^{Z,(r)})_B]^{d_{l,r}^X} + g_{B,2}^{(r)} \Big| + \Big| \sum_{r=1}^{n-1} \big( \sum_l  [(\partial L_{R,l}^{Z,(r)})_B]^{d_{l,r}^X} + g_{B,2}^{(r)} \big) \Big| \Big\}}
     \nonumber \\
     \calZ_{B,3}^{(\mathbf{d}^Z_{l_1}, \mathbf{d}^Z_{l_2})} &= \sum_{ \{f_{B,3}^{(r)}\} } e^{-\mu \Big\{ \sum_{r=1}^{n-1} \Big|\sum_l [(\partial L_{G,l}^{X,(r)})_B]^{d_{l,r}^Z} + g_{B,3}^{(r)} \Big| + \Big| \sum_{r=1}^{n-1} \big( \sum_l [(\partial L_{G,l}^{X,(r)})_B]^{d_{l,r}^Z} + g_{B,3}^{(r)} \big) \Big| \Big\} }
     \nonumber
\end{align}

Thus, the Rényi coherent information is 
    \begin{align}
         I_c^{(n)}(\mathrm{R}, \QM) = -2\log2 + \frac{1}{n-1} \log \Bigg[ & \frac{  \sum_{\mathbf{d}^X_{l_1}, \mathbf{d}^X_{l_2}} \calZ_{R,1}^{(\mathbf{d}^X_{l_1}, \mathbf{d}^X_{l_2})} \calZ_{G,2}^{(\mathbf{d}^X_{l_1}, \mathbf{d}^X_{l_2})} \calZ_{B,2}^{(\mathbf{d}^X_{l_1}, \mathbf{d}^X_{l_2})} }{\calZ_{R,1} \calZ_{G,2} \calZ_{B,2}}
         \nonumber \\
         &\frac{\sum_{\mathbf{d}^Z_{l_1}, \mathbf{d}^Z_{l_2}} \calZ_{R,2}^{(\mathbf{d}^Z_{l_1}, \mathbf{d}^Z_{l_2})} \calZ_{G,1}^{(\mathbf{d}^Z_{l_1}, \mathbf{d}^Z_{l_2})} \calZ_{B,1}^{(\mathbf{d}^Z_{l_1}, \mathbf{d}^Z_{l_2})}  \calZ_{B,3}^{(\mathbf{d}^Z_{l_1}, \mathbf{d}^Z_{l_2})}  }{\calZ_{R,2} \calZ_{G,1} \calZ_{B,1} \calZ_{B,3}} \Bigg].
    \end{align}
The free energy cost of inserting defects along a homologically nontrivial loop is 
\begin{align}
    \Delta F_{\mathbf{d}_{l_1}^X,\mathbf{d}_{l_2}^X} &\equiv -\log \frac{   \calZ_{R,1}^{(\mathbf{d}^X_{l_1}, \mathbf{d}^X_{l_2})} \calZ_{G,2}^{(\mathbf{d}^X_{l_1}, \mathbf{d}^X_{l_2})} \calZ_{B,2}^{(\mathbf{d}^X_{l_1}, \mathbf{d}^X_{l_2})}  }{\calZ_{R,1} \calZ_{G,2} \calZ_{B,2}},
    \quad \Delta F_{\mathbf{d}_{l_1}^Z,\mathbf{d}_{l_2}^Z} \equiv -\log \frac{\calZ_{R,2}^{(\mathbf{d}^Z_{l_1}, \mathbf{d}^Z_{l_2})} \calZ_{G,1}^{(\mathbf{d}^Z_{l_1}, \mathbf{d}^Z_{l_2})} \calZ_{B,1}^{(\mathbf{d}^Z_{l_1}, \mathbf{d}^Z_{l_2})} \calZ_{B,3}^{(\mathbf{d}^Z_{l_1}, \mathbf{d}^Z_{l_2})} }{\calZ_{R,2} \calZ_{G,1} \calZ_{B,1} \calZ_{B,3}}.
\end{align}
We note that in this specific calculation, the stat-mech model $\calZ_{B,2}$ undergoes the ferromagnetic transition first because it has the largest coupling $6\mu$. 
As a result, $\Delta F_{\mathbf{d}_{l_1}^X,\mathbf{d}_{l_2}^X}$ diverges before $\Delta F_{\mathbf{d}_{l_1}^Z,\mathbf{d}_{l_2}^Z}$, and the coherent information undergoes a transition from $2\log 2$ to $0$ and then to $-2\log 2$.
However, this is because we only evolve the state for four rounds in the measurement sequence when computing the coherent information.
If we had considered the evolution longer, the coherent information would undergo a single transition from $2\log 2$ to $-2\log 2$.

\end{document}